
\magnification=\magstep1
\baselineskip=15pt
\overfullrule=0pt

\def\a{\alpha}
\def\ax{{\alpha\cdot x}}
\def\ra{{l^\vee (\alpha)}}
\def\dv{\delta^{\vee}}
\def\o{\omega}
\def\l{{\rm log\,}}
\def\T{\vartheta_1}
\def\s{{\rm sinh\,}}

\def\C{{\bf C}}

\def\G{{\cal G}}
\def\H{{\cal H}}
\def\K{{\cal K}}

\def\N{{\cal N}}

\def\R{{\cal R}}
\def\half{{1 \over 2}}
\def\12{{\scriptstyle{1\over 2}}}
\def\31{{\scriptstyle{1\over 3}}}
\def\23{{\scriptstyle{2\over 3}}}
\def\43{{\scriptstyle{4\over 3}}}
\def\quarter{{\scriptstyle{1 \over 4}}}
\def\Re{{\rm Re}}

\def\f27{{{\bf 26 }$\oplus${\bf 1}}}

\rightline{UCLA/98/TEP/10}
\rightline{Columbia/Math/98}
\rightline{NSF-ITP-98-061}

\bigskip

\centerline{{\bf CALOGERO-MOSER AND TODA SYSTEMS FOR TWISTED
AND}}

\medskip

\centerline{{\bf UNTWISTED AFFINE LIE ALGEBRAS}
\footnote*{Research supported in part by
the National Science Foundation under grants PHY-95-31023,
PHY-94-07194 and DMS-95-05399.}}

\bigskip
\bigskip
\bigskip

\centerline{{\bf Eric D'Hoker}${}^1$ 
            {\bf and D.H. Phong} ${}^2$}

\bigskip

\centerline{${}^1$ Department of Physics}
\centerline{University of California, Los Angeles, CA 90024, USA;}
\centerline{Institute for Theoretical Physics}
\centerline{University of California, Santa Barbara, CA 93106, USA}

\bigskip

\centerline{${}^2$ Department of Mathematics}
\centerline{Columbia University, New York, NY 10027, USA}

\bigskip
\bigskip
\bigskip

\centerline{\bf ABSTRACT}

\bigskip

The elliptic Calogero-Moser Hamiltonian and Lax pair associated with a general
simple Lie algebra $\G$ are shown to scale to the (affine) Toda Hamiltonian and
Lax pair. The limit consists in taking the elliptic modulus $\tau$ and the
Calogero-Moser couplings $m$  to infinity, while keeping fixed the combination 
$M = m e^{i \pi \delta \tau}$ for some exponent $\delta$.  Critical scaling 
limits arise when $1/\delta$ equals the Coxeter number or the dual Coxeter 
number for the untwisted and twisted Calogero-Moser systems respectively; the 
limit consists then of the Toda system for the affine Lie algebras $\G^{(1)}$ 
and $(\G ^{(1)})^\vee$. The limits of the untwisted or twisted
Calogero-Moser system, for $\delta$ less than these critical values, but
non-zero, consists of the ordinary Toda system, while for $\delta =0$, it 
consists of the trigonometric Calogero-Moser systems for the algebras $\G$ and 
$\G^\vee$ respectively.

\bigskip

\vfill\break

\centerline{\bf I. INTRODUCTION}

\bigskip

It is now well recognized that the low energy dynamics of four-dimensional
${\cal N}=2$ supersymmetric gauge theories are governed effectively
by integrable models. While it is not yet known
which models arise in this manner, the models
defined by Lie algebras are naturally expected to play a major role.

\medskip

This paper is the second of a series [1] devoted to the study of twisted
and untwisted elliptic Calogero-Moser systems defined by general simple Lie
algebras, and of their role in Seiberg-Witten theory. In [2], on the basis
of several consistency checks, Donagi and Witten
had proposed that the low energy dynamics of the
$SU(N)$ gauge theory with matter in the adjoint
representation was described by a $SU(N)$ Hitchin 
systems.\footnote{${}^\dagger$}{Extensive references to research on the 
connections between integrable models and supersymmetric Yang-Mills theory may 
be found in [1]. Further references to the derivation of Seiberg-Witten curves 
from effective field theories emerging on branes in string theory and M-theory, 
as well as from singularities in Calabi-Yau compactifications are in [8].}
This was verified in [3] by evaluating explicitly the prepotential, using the 
identification [4][5] of the $SU(N)$ Hitchin system with the $SU(N)$ elliptic 
Calogero-Moser system.
The elliptic Calogero-Moser system is associated with an elliptic curve 
$\Sigma$ (or torus), defined in terms of the periods $2\omega _1$ and $2 
\omega_2$ by $\Sigma \equiv {\bf C}/(2\o_1{\bf Z}+2\o_2{\bf Z})$.
The modulus $\tau=\o_2/\o_1$ of $\Sigma$ is related to the gauge coupling
of the super-Yang-Mills theory by
$$
\tau = {4 \pi i \over g^2} + {\theta \over 2 \pi }.
\eqno (1.1)
$$
The Calogero-Moser Lax pair of operators $L(z)$, $M(z)$ depend on an arbitrary   
spectral parameter $z\in \Sigma$, and the Lax equation
$$
\dot L(z)=[L(z),M(z)]
\eqno (1.2)
$$ 
is equivalent to the Hamilton-Jacobi equations of the elliptic Calogero-Moser 
system. In terms of these data, the Seiberg-Witten curve is precisely the
Calogero-Moser spectral curve 
$$
\Gamma=\{(k,z); \ \det(kI-L(z))=0\},
\eqno (1.3)
$$
and the Seiberg-Witten differential is $d\lambda=kdz$. The theme of this series 
of papers is to extend this analysis to gauge theories associated with an 
arbitrary simple Lie algebra $\G$. 

\medskip

In the first paper of the series [1], we had indicated that
besides the usual elliptic Calogero-Moser systems defined by Lie algebras,
the extension to non-simply laced algebras actually required
the introduction of new systems, namely the
{\it twisted} elliptic Calogero-Moser systems.
We had also constructed explicitly Lax pairs with spectral parameters for
all (twisted and untwisted) elliptic Calogero-Moser systems, except in the case 
of twisted $G_2$. Now the identification of which
integrable model corresponds to any given gauge theory is still 
largely conjectural, and no direct derivation is available thus far.
Rather, as in the original case of the
$SU(N)$ four-dimensional gauge theory and Hitchin systems studied by
Donagi and Witten, the identification
of the correct integrable model is usually based on consistency checks
such as limits of the theories as mass parameters tend to 0 or infinity.
The goal of the present paper is to describe these limits
and consistency checks in the case of elliptic Calogero-Moser systems, 
and explain why the twisted models
are required for non-simply laced algebras. 

\medskip

It is well-known (see Inozemtsev [6-7]) that the elliptic Calogero-Moser
system corresponding to $\G=A_n=SU(n+1)$
$$
H=\half\sum_{i=1}^{n+1}p_i^2-\half m^2\sum_{i\not=j}\wp(x_i-x_j)
\eqno (1.4)
$$
tends to either the Toda or the periodic Toda system 
$$
\eqalign{
H&=
\half\sum_{i=1}^{n+1}P_i^2-\half\sum_{i=1}^{n}e^{X_{i+1}-X_i}, \ \ 
({\rm Toda})\cr
H&=
\half\sum_{i=1}^{n+1}P_i^2-\half\big(\sum_{i=1}^{n}e^{X_{i+1}-X_i}+
e^{X_1-X_{n+1}}\big), \ \ ({\rm periodic\ Toda})\cr}
\eqno (1.5)
$$
in the limit where $\o_1=-i\pi$, $\o_2\rightarrow\infty$, and
$$
\eqalign{
x_i&=X_i-2\o_2\delta \, i,\ \ \ p_i=P_i,\ \ \ 1\leq i\leq n+1,\cr
m&=Me^{-i \pi \delta \tau},\cr}
\eqno(1.6)
$$ 
depending on whether $\delta< 1/(n+1)$ or $\delta= 1/(n+1)$.
(Other limits are also discussed in [6],
but we do not need them here). We shall be mainly interested in 
extensions of the
critical case $\delta=1/(n+1)$, although the subcritical case
is easily treated by the same arguments.

\medskip

{}For general Lie algebra $\G$, the scaling prescription (1.6) admits two 
distinct generalizations, depending essentially on whether the critical value 
$1/\delta=n+1$ is replaced by the {\it Coxeter number}\footnote{*}{Some key 
facts about Lie algebra theory, and in particular about the Coxeter numbers and 
dual Coxeter numbers are given in the Appendix \S A of [1].} $h_{\G}$ or by the 
{\it dual Coxeter number} $h_{\G}^{\vee}$. 
For our purposes, $h_{\G}$ and
$h_{\G}^{\vee}$ are most conveniently defined in the following manner. Let 
$\a_i$, 
$1\leq i\leq n$,
be a basis of simple roots for the Lie algebra $\G$. For each root $\a$,
the coroot $\a^{\vee}$ is defined by $\a^{\vee}={2\a\over\a^2}$.
Now expand $\a$ and $\a^{\vee}$ respectively in terms of the bases
$\{\a_i\}$ of simple roots and $\{\a_i^{\vee}\}$ of simple coroots
$$
\a=\sum_{i=1}^{n} l_i\a_i,\ \qquad
\a^{\vee}=\sum_{i=1}^{n} l_i^{\vee}\a_i^{\vee}.
\eqno (1.7)
$$
Then $h_{\G}$ and $h_{\G}^{\vee}$ are defined as the following maxima when the 
root $\alpha$ runs through the root system of $\G$,
$$
h_{\G}=1+{\rm max} \sum_{i=1}^{n}l_i,\ \qquad   
h_{\G}^{\vee}=1+{\rm max}\sum_{i=1}^{n}l_i^{\vee}.
\eqno (1.8)
$$   
The Coxeter and dual Coxeter numbers are evidently the same when $\G$ is simply
laced, but otherwise $h_{\G}^{\vee}$ is strictly less than $h_{\G}$.
Now it is not difficult to show that non-trivial limits can only arise 
when the dynamical variable $x$ scales according to $x=X+(2\omega_2)v$ for 
some fixed vector $v$ in 
${\bf R}^n$ (c.f. \S II below). Depending on whether
$1/\delta$ is $h_{\G}$ or $h_{\G}^{\vee}$ (or equivalently, on whether we want 
the 
simple
roots of $\G^{(1)}$ or of $(\G^{(1)})^{\vee}$ to survive in the limits),
we have to make the following choices for the vector $v$

\medskip
$\bullet$ $x=X+2\o_2\delta\rho^{\vee}$, if $m=Me^{-i \pi \delta \tau}$,
$\delta\leq 1/h_{\G}$;

$\bullet$ $x=X+2\o_2\delta^{\vee}\rho$, if $m=Me^{-i \pi \delta^{\vee} \tau}$,
$\delta^{\vee}\leq 1/ h_{\G}^{\vee}$.

\medskip
\noindent
Here $\rho$ and $\rho^{\vee}$ are respectively the Weyl vector
and the level vector.

\medskip

More precisely, we shall show that, under the scaling rules associated
with $\delta=1/h_{\G}$,
the {\it untwisted} elliptic Calogero-Moser Hamiltonian and its Lax pair
with spectral parameter recently constructed in [1] converge
to the affine Toda Hamiltonian and Lax pair associated with
$\G^{(1)}$. For $\delta$
less than this critical value, but non-zero, the limit consists of the ordinary
Toda system for $\G$, while for $\delta =0$, we find the trigonometric
Calogero-Moser system for $\G$. Under the scaling rules associated with
$\delta=1/h_{\G}^{\vee}$, when $\G$ is not simply laced,
the untwisted elliptic Calogero-Moser systems do not converge
to a finite limit. However,  
the new {\it twisted} elliptic Calogero-Moser
systems introduced in [1] as well as all the
Lax pairs constructed there do converge, and the limit consists
of the affine Toda system for the affine Lie algebra $(\G ^{(1)})^\vee$. 
For $\delta$ less than this critical value, but non-zero, the limit consists in 
the
ordinary Calogero-Moser system for $\G ^\vee$, while for $\delta =0$, we find
the trigonometric Calogero-Moser system for $\G ^\vee$.

\medskip

We had mentioned earlier that
the scaling limits in this paper constitute a key piece of
evidence for identifying the integrable model describing
the low energy effective theory of the ${\cal N}=2$ supersymmetric
$\G$ gauge theory with a hypermultiplet in the adjoint representation of $\G$.
A detailed discussion together with some of the underlying physics
is postponed to the third paper of this series [8]. Here we note only that
our results strongly suggest that the {\it twisted} Calogero-Moser
systems associated with $\G$ are the correct integrable models.
The mass of the adjoint
hypermultiplet is given by the Calogero-Moser coupling constant. For simply
laced $\G$, there is just one such coupling $m$, which is the hypermultiplet
mass. For non-simply laced $\G$, there are two such Calogero-Moser couplings, 
one
for long and one for short roots, $m_l$ and $m_s$ respectively, and both are
proportional to the mass $m$ with known group theoretical factors. The
limits established here correspond to letting $m \to \infty$ and thus to
decoupling the hypermultiplet. The key issue is which scaling rule
is the appropriate rule dictated by physics. Identifying the dual Coxeter 
number
$h_{\G}^{\vee}$ with the quadratic Casimir of $\G$, 
the dependence of the gauge coupling on the mass $m$ in this limit is given by
$$
\tau = {i \over 2 \pi} h_\G ^\vee\, {\rm ln} \,{m ^2  \over M^2},
\eqno (1.9)
$$
in view of standard renormalization group arguments.
Thus the scaling rules associated with the dual Coxeter number $h_\G ^\vee $ 
are the appropriate ones, and with them, the twisted elliptic
Calogero-Moser systems. As $m\to 0$ and $m\to\infty$, the desired limits  
emerge

\medskip
\item{(1)} At $m=0$, the integrable model is free, corresponding to the
fact that the gauge theory acquires an ${\cal N}=4$ supersymmetry,
and the prepotential receives no quantum corrections. 
\item{(2)} At $m=\infty$, the limit of the Calogero-Moser system is a 
twisted affine Toda system, which was previously argued by Martinec and Warner 
[9] to be associated with the $\N=2$ supersymmetric
gauge theory without hypermultiplets.

\medskip

{}Further evidence may be obtained by comparing the prepotential 
derived in the 
weak coupling limit, $\tau \to +i\infty$, of the twisted elliptic 
Calogero-Moser systems with that predicted by one-loop calculations
in the gauge theory. These calculations may be carried out using the explicit 
form of the Lax pairs in [1], and using the methods of [3] and [20]. As an 
example, this check is carried out successfully for $\G=D_n$ in [8].

\bigskip

\noindent
{\bf Calogero-Moser and Toda Systems and their Interrelation}

\medskip

Let $\G$  be a simple finite-dimensional Lie algebra of rank $n$, and denote 
the set of all roots of $\G$ by $\R (\G)$.
The Toda and Calogero-Moser systems are Hamiltonian systems with $n$ complex
degrees of freedom and their canonical momenta, denoted by $X_i$ and $P_i$ for
Toda and by $x_i$ and $p_i$ for Calogero-Moser,\footnote*{In the case of $A_n$, 
as we saw in the Introduction, it is sometimes more convenient to have $n+1$ 
dynamical variables variables $(X_i,P_i)$ or $(x_i,p_i)$.
The correct rank $n$ is restored upon observing that the dynamical variables
can be shifted by an arbitrary constant.} with $i=1,\cdots ,n$. We assemble 
these degrees of freedom into $n$-dimensional vectors $X$, $P$, $x$ and $p$, 
and use the dot notation for inner products.

\medskip

The Toda system associated with a finite-dimensional 
or affine Lie algebra $\K$ 
is defined by the Hamiltonian 
$$
H_T
=
 \half P\cdot P -\half \sum _{\alpha \in \R_* (\K)}
  M_{|\alpha|} ^2 e^{- \alpha \cdot X},
\eqno (1.10)
$$
where $M _{|\alpha|}$ are constants and $R_* (\K)$ is the set of 
simple roots of $\K$. 

\noindent
$\bullet$ When $\K=\G$ is any finite-dimensional Lie algebra, the system $H_T$ 
is referred to as the ordinary (or non-periodic) Toda system associated with 
$\G$.  

\noindent
$\bullet$ When $\K$ is any of the affine Lie algebras, the system $H_T$ is 
referred to as the affine (or periodic) Toda system associated with $\K$. 

\medskip

\medskip

The (untwisted) elliptic Calogero-Moser (CM) system is defined for any finite 
dimensional Lie algebra $\G$  by the Hamiltonian
$$
H_{CM} =   
 \half p \cdot p -\half\sum _{\alpha \in \R(\G)} m_{|\alpha|} ^2
\wp (\alpha \cdot x),
\eqno  (1.11) 
$$
where $\wp $ is the Weierstrass elliptic function of periods $2 \omega _1$ and
$2 \omega _2$ of the underlying elliptic curve $\Sigma$. 

\medskip

The twisted Calogero-Moser systems (TCM) may be defined 
for any finite-dimensional Lie algebra $\G$ by the Hamiltonian
$$
H_{TCM} =   
\half p\cdot p 
-\half \sum _{\alpha \in \R(\G )} 
 m_{|\alpha|}^2 \wp _{\nu (\alpha)} (\alpha \cdot x). 
\eqno (1.12) 
$$
For simply laced $\G$, we have $\nu =1$ on all roots and the twisted 
Calogero-Moser system is identical to the untwisted one of (1.11). 
Henceforth, we shall assume that $\G$ is non-simply laced. The root system is
then a union  of the set of long roots $\R_l$ and the set of short roots 
$\R_s$.
On long roots, $\nu =1$, while on short roots $\nu (\alpha)$ 
equals the ratio of the length squared of the long roots to the short roots.
Thus, $\nu=2$ for 
$\G = B_n$, $C_n$, 
$F_4$, while $\nu=3$ for $\G=G _2$. The 
function $\wp _\nu$ is twisted of order $\nu$ in one of the
three half periods $\omega _1$, $\omega _2$ or $\omega _3 = \omega _1 + \omega
_2$ :
$$
\wp _\nu (u) = \sum _{\sigma =0} ^{\nu -1} \wp (u + 2 \omega _a {\sigma \over
\nu}).
\eqno (1.13)
$$
In the sequel, it will be convenient to choose $\omega _a = \omega _1$. The
trigonometric and rational Calogero-Moser systems are obtained from the 
(untwisted) elliptic
systems (for each Lie algebra) by letting respectively one or both of the
periods $2 \omega _1$ and $2\omega _2$ tend to infinity
$$
\eqalign{
H_{CM}^{trig}
=& \half p \cdot p-\half\sum _{\alpha \in \R(\G )} 
 m_{|\alpha|}^2{1\over{\rm sinh}^2\,(\ax)}\cr
H_{CM}^{rat}
=& \half p \cdot p -\half\sum _{\alpha \in \R(\G )} 
 m_{|\alpha|}^2{1\over (\ax)^2}.\cr}
 \eqno (1.14)
$$

\medskip

We shall now summarize the results of this paper in the form of the Theorems 1
and 2 below.\footnote{$\dagger$}{Henceforth, we shall set $\omega _1 = -i\pi$, 
so that $\tau = i \omega _2 /\pi$, and $q = e^{2 \pi i \tau} = e^{-2 \omega 
_2}$.} Recall that $h_\G$ and $h^\vee _\G$ denote respectively the Coxeter 
and dual Coxeter numbers of the finite-dimensional Lie algebra $\G$, 
$\rho$ the Weyl vector, and $\rho ^\vee$ the level vector of $\G$. 
When $\Re(\omega _2) \to \infty$ and $m_{|\alpha |} \to \infty$, while keeping
the quantities $M_{|\alpha|}$, $X$ and $P$ fixed, we have the limits below.

\bigskip

\noindent
{\bf Theorem 1 : The Untwisted Cases} 

The scaling behavior is governed by an exponent $\delta$ and is given by
$$
\eqalignno{
M _{|\alpha | } = & \ m_{|\alpha |} q^{\half\delta},  
& (1.15a) \cr
X = & \ x  - 2 \omega _2\, \delta \,\rho ^\vee ,
\qquad \qquad P= p,
&   (1.15b) \cr
Z = & \ e^z\,e^{- i \pi \tau }. &(1.15c)
\cr }
$$
The Hamiltonian $H_{CM}$ of the {\it untwisted} elliptic
Calogero-Moser system  for the Lie algebra $\G$, converge to those of the
\item{(a)} affine (periodic) Toda system with {\it untwisted} affine Lie
algebra $\G ^{(1)}$ when $\delta = 1/h_\G$;
\item{(b)} ordinary (non-periodic) Toda system with Lie algebra $\G$ when
$\delta < 1/h_\G$;
\item{(c)} trigonometric Calogero-Moser system with Lie algebra $\G$ when
$\delta =0$.

The Lax pairs constructed in [1] for all untwisted, elliptic Calogero-Moser
systems defined by simple Lie algebras converge to Lax pairs for the 
corresponding affine Toda system for $\G^{(1)}$ 
(when $\delta=1/h_{\G}$), Toda system for $\G$ (when $\delta<1/h_{\G}$),
and trigonometric Calogero-Moser system for $\G$ (when $\delta=0$). (The 
case of $E_8$ was solved in [1] making use of an extra assumption on the 
existence of a $\pm 1$-values cocycle. The same assumption is implied here.)

\bigskip

\noindent
{\bf Theorem 2 : The Twisted Cases}

The scaling behavior is governed by an exponent $\delta ^\vee$ and is given by
$$
\eqalignno{
M _{|\alpha | } = & \ m_{|\alpha |} q^{\half\delta^{\vee}},  
& (1.16a) \cr
X = & \ x  - 2 \omega _2\, \delta^{\vee} \,\rho  ,
\qquad \qquad P= p,
&   (1.16b) \cr
Z = & \ e^z\,e^{-i \pi \tau}. &(1.16c)
\cr }
$$
The Hamiltonian $H_{TCM}$ of the {\it twisted} elliptic
Calogero-Moser system associated with a Lie algebra $\G$, converge to those of 
\item{(a)} affine (periodic) Toda system with {\it twisted} affine Lie
algebra $(\G ^{(1)}) ^\vee$ when $\delta ^\vee = 1/h_\G ^\vee $.
\item{(b)} ordinary (non-periodic) Toda system with Lie algebra $\G ^\vee $
when $\delta ^\vee < 1/h_\G ^\vee $;
\item{(c)} trigonometric Calogero-Moser system with Lie algebra $\G ^\vee $
when $\delta  ^\vee =0$.

The Lax pairs constructed in [1] for all twisted, elliptic Calogero-Moser
systems defined by simple Lie algebras except $G_2$ converge to Lax pairs
of the corresponding affine Toda system for $(\G^{(1)})^{\vee}$
(when $\delta^{\vee}=1/h_{\G}^{\vee}$), Toda system for $\G^{\vee}$ 
(when $\delta^{\vee}<1/h_{\G}$), and trigonometric
Calogero-Moser system for $\G^{\vee}$ (when $\delta^{\vee}=0)$.  

\bigskip
The Lax pairs of the Toda and Calogero-Moser systems will be presented
explicitly in the subsequent sections of this paper.

\bigskip

The remainder of this paper is devoted to a complete proof of Theorems 1 and 2.
In \S II, we discuss and prove Theorem 1 : the limits of the
Hamiltonians and Lax pairs for the {\it untwisted} Calogero-Moser systems.
We discuss and prove Theorem 2 on the limits of the {\it twisted}
Calogero-Moser systems in \S III for the Hamiltonians and in \S IV for the Lax
pairs. For a discussion of the Lie algebra and elliptic function
results we need, we refer the reader to Appendices \S A and \S B of 
[1] respectively. Other useful references on Lie algebras are [10-12].
Surveys of earlier work on integrable models associated with Lie
algebras can be found in [13].

\bigskip
\bigskip

\centerline{\bf II. UNTWISTED CALOGERO-MOSER AND (AFFINE) TODA SYSTEMS}

\bigskip

{}First we recall the expressions for Weierstrass elliptic functions in terms
of Jacobi theta functions, as well as their product and series expansions which
will be useful when considering the limit $\Re (\omega _2) \to \infty$.
It is convenient to introduce the following modification of the standard $\T$ 
function, with its product expansion
$$
\T^*(u|\tau) 
\equiv  2\pi i {\T({u \over 2 \pi i}|\tau)\over\T'(0|\tau)}
=  
2\,{\rm sinh}\,({u \over 2})
\prod_{n=1}^{\infty}
(1-q^ne^{u})(1-q^ne^{-u})(1-q^n)^{-2}.
\eqno (2.1)
$$
Then the Weierstrass functions $\sigma(u)$, $\zeta(u)$, and $\wp(u)$
are defined by [14] 
$$
\eqalignno
{\sigma(u)=&e^{-{\eta_1\over 2\pi i}u^2}\T^*(u|\tau)\cr
\zeta(u)=&-{\eta_1\over \pi i}u+\partial_u\l\T^*(u|\tau)\cr
\wp(u)=& {\eta_1\over \pi i}-\partial_u^2\l\T^*(u|\tau).
&(2.2)\cr}
$$
It will be very convenient to express the elliptic function $\wp$ as a series
expansion involving hyperbolic functions [6]
$$
\wp(u)
={\eta_1\over\pi i}
+\half \sum_{n=-\infty}^{\infty}
{1\over \cosh (u-2n\omega _2) -1 }.
\eqno (2.3)
$$
The series expansion has the advantage of being uniformly convergent throughout
$u \in \C$, as long as $\Re (\omega _2 )>0$.  The constant $\eta_1 = \zeta
(\omega _1)$ may be determined from the fact that
$\wp(u)=u^{-2}+O(u^2)$. Henceforth, we shall neglect it since it does not 
affect
the Hamilton-Jacobi equations of the systems.

\bigskip

\noindent
{\bf A. The Scaling Limit of the Hamiltonian}

\medskip

Our first task is to derive the limit of the Hamiltonian of the Calogero-Moser
system as $\Re (\omega _2)\to \infty$, keeping $M_{|\alpha|}$ fixed,
according to 
$$
m_{|\alpha |}
=M_{|\alpha |} e^{\delta \omega _2}.
\eqno (2.4)
$$
Here $\delta$ is a real scaling exponent with $\delta \geq 0 $, to be 
determined
later. It suffices to take the above limit 
of the combination $m_{|\alpha|} ^2 \wp (\alpha \cdot x)$
separately for each root $\alpha \in \R(\G)$.
Using the
series representation for $\wp $ of (2.3), we have 
$$
m_{|\alpha|} ^2 \wp (\alpha \cdot x)
=
\half \sum _{n=-\infty} ^ \infty 
{M_{|\alpha |}^2  e^{2 \delta \omega _2} 
\over 
\cosh (\alpha \cdot x -2n\omega _2) -1}. 
\eqno (2.5)
$$ 
Since this series is uniformly convergent throughout $\C$, we may analyze its
limit term by term in (2.5). Clearly, if $\delta =0$, only the term $n=0$ will
survive in the limit, and we recover the trigonometic Calogero-Moser system,
which proves (c) of Theorem 1 for the Hamiltonian. 

\medskip

Henceforth, we assume that $\delta >0$, so that $m_{|\alpha |} \to \infty$ as
$\Re (\omega _2) \to \infty$. We begin by giving a justification for the 
scaling
behavior announced in (1.15b). The condition $P=p$ is manifest. It is clear 
from
(2.5) that unless $ x$ has a non-trivial dependence on $\omega _2$, the limit 
of
$m_{|\alpha|} ^2 \wp  (\alpha \cdot x)$ will diverge. It follows from
the form of (2.5) that  the only interesting $\omega _2$ dependence of
$x$ is by a shift linear  in $\omega _2$. Thus, we set $x=X + 2 \omega _2 v,$
for some vector $v$  in ${\bf R} ^n$, and keep $X$ and $v$ fixed as $\Re
(\omega _2) \to \infty$. 

\medskip

A number of constraints on the vector $v$ result from the following 
considerations. The $n=0$ term in (2.5) will diverge unless $|v \cdot
\alpha| \geq \delta $ for all roots $\alpha$. 
To analyze this constraint in more detail, we fix a basis of simple
roots $\a_i$, $1\leq i\leq n$ for $\G$. Then any root $\a$
of $\G$ may be written as
$$
\a=\sum_{i=1}^n l_i\a_i,
\eqno (2.6)
$$  
with all $l_i\geq 0$ for positive roots, and all $l_i\leq 0$ for negative 
roots.
A finite limit of (2.5) requires that $|v\cdot \alpha _i| \geq \delta$ for all
simple roots $\alpha _i$. Without loss of generality, we may assume that all
inner products are positive, so that the constraint becomes $v \cdot \alpha
_i \geq \delta$. Once this holds, the constraint $|v\cdot \alpha | \geq
\delta$ will be  satisfied for all roots $\alpha$ in view of (2.6). The 
situation
where the inequality is  saturated for every simple root produces the maximal
number of roots  surviving in the limit, and will thus result in a {\it 
maximally
symmetric  limit}. All other cases can be reduced to those upon considering
directly the  Calogero-Moser system associated with a subalgebra of
$\G$. Henceforth, we shall only consider the {\it maximally
symmetric limits}. 

\medskip

In any finite dimensional simple Lie algebra, there exists a unique vector, 
whose inner product with any simple root is 1. This is the {\it level vector} 
$\rho ^\vee$, defined as the half sum of all positive coroots. 
The inner product of $\rho^\vee $ with the root $\alpha$ (and more generally 
with any weight of $\G$), defines the {\it level} \ function $l(\alpha)$ by
$$
l(\alpha ) \equiv \alpha \cdot  \rho ^\vee.  
\eqno (2.7)
$$  
For any $\G$, and any root $\alpha$, the level $l(\alpha)$ is an integer,
and takes the value 1 if and only if $\alpha$ is a simple root. 
It is clear then that the {\it maximally symmetric limits} correspond to
$v$ proportional to the {\it level vector} $\rho ^\vee$, with proportionality
factor $\delta$. We thus recover (1.15b), or equivalently, using (2.7)
$$
\alpha \cdot x =  \alpha \cdot X + 2 \omega _2 \delta \,  l(\alpha). 
\eqno (2.8)
$$
The limit of the $n=0$ term is now finite under the scaling (1.15b) or (2.8).

\medskip

In order that contributions to (2.5) for all $n$ have finite limits, further
constraints must be imposed.  By periodicity of $\wp$, it is easy to see that 
the
product $\delta \,  l(\alpha)$ must stay away from any integer value by a
distance of at least $\delta$. In other words, we must have 
$$
0<\delta \leq \delta \, l(\alpha ) - [ \delta \, l(\alpha)] \leq 1-\delta,
$$
where $[a]$ is the integer part of $a$. The simplest way to realize this extra 
constraint is to require that $\delta \, l(\alpha ) <1$ for all positive roots.
This will be the case throughout the paper. Then the
preceding condition becomes
$$\delta \leq \delta \, l(\alpha) \leq 1-\delta
\eqno(2.9)
$$ for all positive roots
$\alpha$. If $\alpha _0$ is the highest root of $\G$, and $l_0 = l(\alpha _0)$
its level, then it suffices that the above condition be satisfied on $\alpha_0$ 
:
$$
h_\G =  1 + l_0  \leq {1 \over \delta}.
\eqno (2.10)
$$
Here $h_{\G}=1+l_0$ is the Coxeter number of $\G$.
The case where $\delta > 1/h_\G$ is more complicated and will be discussed 
in a forthcoming publication.
The evaluation of the limits below
relies on the fact that, in the critical case where $\delta=1/h_{\G}$,
the first inequality in (2.9) becomes an equality if and only if $\alpha$
is a simple root, while the second inequality in (2.9)
becomes an identity if and only if $\a$ is the highest root $\a_0$.

\medskip

\noindent
{\it General Limit Formulas}

Since $\wp$ is even, it suffices to consider positive roots $\a$. In view of 
the
asymptotics 
$$
2 \{\cosh (\alpha \cdot x - 2n\omega _2) -1 \}
\to \left \{
\matrix{
e^{+\a\cdot X+2\omega_2(\delta\, l(\a)-n)}, & {\rm if}\ \delta\, l(\a)-n>0\cr
& \cr
e^{-\a\cdot X-2\omega_2(\delta\, l(\a)-n)}, & {\rm if}\ \delta\, l(\a)-n<0,
\cr} \right .
\eqno (2.11)
$$
we have the following limiting behavior
$$
m_{|\a|}^2\wp(\ax)
\to M_{|\a|}^2\bigg[
\sum_{n<\delta\, l(\a)}
e^{-2\omega_2(\delta\, l(\a)- n-\delta)-\a\cdot X}
+ 
\sum_{n>\delta\, l(\a)}
e^{-2\omega_2(n-\delta\, l(\a)-\delta) + \a\cdot X}\bigg].
\eqno(2.12)
$$
This expression can be made more explicit upon the assumption that
$0<\delta <\delta \, l(\alpha) <1-\delta<1$, introduced above for positive 
roots
$\alpha$,
$$
m_{|\a|}^2\wp(\ax)
\to M_{|\a|}^2\bigg[
\sum_{n\geq 0}e^{-2\omega_2(n+\delta\, l(\a)-\delta) -\a\cdot X}
+ 
\sum_{n\geq 1}e^{-2\omega_2(n-\delta\, l(\a)-\delta) + \a\cdot X}\bigg].
\eqno (2.13)
$$
In the first sum, all contributions with $n\geq 1$ converge to zero, and may be
ignored in the limit $\Re (\omega _2) \to \infty$. In the second sum, all
contributions with $n \geq 2$ converge to zero and may be ignored as well. We 
are
thus left with the following asymptotics
$$
m_{|\a|}^2\wp(\ax)
\to 
M_{|\a|}^2 \bigg[ 
 e^{-2\omega_2(\delta\, l(\a)-\delta) - \alpha \cdot X}
+e^{-2\omega_2(1-\delta\, l(\a)-\delta) + \alpha \cdot X}
\bigg]
\eqno (2.14)
$$
Depending upon the range of values for $\delta$, this limit produces the 
ordinary
Toda or the affine Toda system. We shall analyze these limits separately.

\bigskip

\noindent
{\it Limit to the ordinary Toda system}

Since $\delta >0$, the limit of the first term in (2.14) is zero for all 
positive roots $\alpha$ for which $l(\alpha ) \geq 2$. Thus, only the
contributions of  the simple roots $\alpha _i$, $i=1,\cdots ,n$ of $\G$ 
survive.
The limit of the second term in (2.14) vanishes for all positive roots $\alpha$
for which $l(\alpha) < 1/\delta -1$. Now, for all positive roots of $\R(\G)$ to
obey this inequality, it suffices that the highest root of $\G$ satisfy the
inequality. But, the level of the highest root of $\G$ is related to the 
Coxeter
number $h _\G$ of $\G$ by $ h_\G = 1 + l _0 $,   
so that the above inequality becomes
$$
\delta < {1 \over h_\G}.
\eqno (2.15)
$$
Thus, whenever $\delta$ satisfies (2.15), the second term on the r.h.s. of
(2.14) will converge to zero for all roots of $\G$. Putting all together, for 
any roots, we have
$$
m_{|\a|}^2\wp(\ax)
\rightarrow
\cases{M_{|\a|}^2e^{\mp \a\cdot X}, & $l(\a)=\pm 1$ \cr
0, & otherwise.\cr}
\eqno (2.16)
$$
The limit of the Hamiltonian $H_{CM}$ for the Lie algebra $\G$ thus yields the
Hamiltonian $H_T$ of the ordinary Toda system for $\G$, as indeed announced in
Theorem 1 (b).

\bigskip

\noindent
{\it Limit to the affine Toda system}

{}From the above discussion, it is clear that the value 
$$
\delta={1\over l_0 +1} = {1 \over h_\G},
\eqno(2.17)
$$
corresponds to a critical case, for which the second term in (2.14) also 
survives
the  limit $\Re (\omega _2) \to \infty$.
We have 
$$
m_{|\a|}^2\wp(\ax)
\rightarrow
M_{|\a|}^2
\cases{e^{\mp \a\cdot X}, & if $l(\a)=\pm 1$;\cr
e^{\pm \a_0\cdot X},& if $l(\a)=\pm l_0$;\cr
0, & otherwise.\cr}
\eqno (2.18)
$$
The limit of the Hamiltonian $H_{CM}$ for the Lie algebra $\G$ then yields the
Hamiltonian $H_T$ of the affine Toda system associated with the
untwisted affine Lie algebra $\G ^{(1)}$, as announced in Theorem 1 (a). Here,
$-\alpha _0$ plays the role of the affine simple  root of $\G ^{(1)}$.

\bigskip

\noindent
{\bf B. The Scaling Limit of the Lax Pair}

\medskip

The Lax operators $L$ and $M$ with spectral parameter $z$, for the (untwisted)
Calogero-Moser systems associated with an arbitrary simple finite dimensional
Lie algebra $\G$ were constructed in [1]. The Lax operators are obtained
starting from an $N$-dimensional representation of $\G$, with weights 
$\{\lambda 
_I\} _{I=1,\cdots , N}$, which embeds $\G$ into $GL(N, \C)$ and are given as
follows
$$
\eqalign{
L=  P+X, 
\qquad & \qquad
P = \sum _{i=1} ^n p_i h_i,  \cr
M= D +Y,
\qquad & \qquad
D = \sum _{i=1} ^n d_i h_i + \sum _{j=n+1} ^N d_j \tilde h _j  + \Delta .
\cr}
\eqno (2.19)
$$
Here, $h_i$, $i=1,\cdots ,n$ generate the Cartan subalgebra $\H _\G$ of $\G$,
$\tilde h_j$, $j=n+1, \cdots, N$ generate the orthogonal complement to $\H_\G$
in the  Cartan algebra of $GL(N, \C)$, and $\Delta$ belongs to the centralizer 
of
$\H_\G$ in $GL(N, \C)$, so that $[D,P]=0$. Finally, $X$ and $Y$ are given by
$$
\eqalign{
X= & \sum _{I,J=1;I\not= J} ^N C_{I,J}  
                             \Phi (\alpha _{IJ} \cdot x, z) E_{IJ} \cr
Y= & \sum _{I,J=1;I\not= J} ^N C_{I,J}  
                             \Phi  ' (\alpha _{IJ} \cdot x, z) E_{IJ}, \cr}
\eqno (2.20)
$$
The combination $\alpha _{IJ} \equiv \lambda _I - \lambda _J$ is the weight
under $\G$ associated with the root $u_I - u_J$ of $GL(N, \C)$, $C_{I,J}$ are
constants, $\Phi  '(x,z)$ is the $x$-derivatives of $\Phi (x,z)$, an elliptic
function that will be defined below. The analysis of [1] implies that the
coefficients $C_{I,J}$ vanish unless $\alpha _{IJ}$ is a root of $\G$, in which
case they are proportional to $m_{|\alpha|}$, and scale in the same way as
$m_{|\alpha|}$ in (2.4),
$$
C_{I,J} =
\left \{ 
\matrix{ M_{|\alpha|}\,e^{\delta \omega _2} c_{I,J} 
& \qquad {\rm when} \quad \alpha _{IJ} = \alpha \in \R (\G) \cr
& \cr 
0
& {\rm when} \quad \alpha _{IJ}  \notin \R (\G). \cr} 
\right .
\eqno (2.21)
$$
Here, the coefficients $c_{I,J}$ are purely group theoretical and were obtained
in [1].

\medskip

To construct a finite limit of the Lax pair $L,M$, we need to make
the spectral parameter $z$ be dependent on $\omega _2$ as well. This is no
problem, since the Lax operators reproduce the Calogero-Moser system for all
values of $z$. The scaling limit indicated in Theorem 1 
$$
e^z=Ze^{-\omega_2}
\eqno(2.22)
$$
where $Z$ is held fixed,
is the limit which generalizes 
the discussion for the special case of the algebra
$A_n$ treated in [3]. Since the Lax pair $L,M$
has been expressed entirely in terms of $m_{|\a|}\Phi(\ax,z)$
and $m_{|\a|}\Phi'(\ax,z)$, the evaluation of their limits
reduces to the evaluation of the limits of 
$m_{|\a|}\Phi(\ax,z)$ and $m_{|\a|}\Phi'(\ax,z)$.
The definition of $\Phi(u,z)$ in terms of $\sigma (z)$ and its expression in 
terms 
of
$\T^*$-functions, are given by [15]
$$ 
\Phi(u,z)=
 {\sigma (z-u) \over \sigma (z) \sigma (u) } e^{u \zeta (z)} 
= 
{\T^*(z-u|\tau)\over
\T^*(z|\tau)\T^*(u |\tau)}
e^{u\partial_z\l\T^*(z|\tau)}.
\eqno(2.23)
$$
To evaluate the asymptotic behavior of this function, we use the
product representation of (2.1). For $z$ satisfying (2.22), the right hand side
of (2.1) can be replaced by $2\,\s{z\over 2}$.
Also  in the limit of interest to us, $u$ in (2.23) is replaced by
$\ax=\alpha\cdot X+ 2\omega_2\,\delta\, l(\alpha)$, with $\delta\,
|l(\alpha)|\leq \delta l_0\leq 1-\delta$. 
Thus a similar approximation is valid for $\T^*(u|\tau)$, and we have
$$
\Phi(u,z)=
\,e^{\12 u {\rm coth\,}{z\over 2}}
\ {\T^*(z-u|\tau)\over
4\, \s{z\over 2}\,\s {u\over 2}}.
\eqno(2.24)
$$
Combining the scaling limits of $x$ and $z$, we have 
$$
z-u=
z-\ax=
-\a\cdot X - \l Z -\omega_2( 1 + 2\delta\, l(\alpha)).
$$ 
The coefficient $1 + 2\delta \, l(\alpha)$ obeys $-1<1 + 2\delta \,
l(\alpha)<3$. Within this range, it suffices to retain the
following asymptotic behavior of $\T^*(z-u|\tau)$ for our purposes,
$$
\T^*(z-u|\tau)
\to 2\,\s {z-u\over 2} (1  - e^{-2 \omega _2 -z + u}),
$$ 
which results in the following asymptotic behavior for $\Phi(u,z)$
$$
\Phi (u,z) \to \left \{
\matrix{
+e^{-\12 u} (1 - Z^{-1} e ^{u- \omega _2} ) & \qquad \Re (u) \to +\infty \cr
&\cr
-e^{+\12 u} (1 - Z e^{-u -\omega _2} )      & \qquad \Re (u) \to - \infty \cr}
\right .
\eqno (2.25)
$$
As the function $\Phi (x,z)$ is not symmetric under $x \to -x$, we
treat the cases of positive and negative roots separately.
 
\medskip
\noindent
{\it Positive Roots}

The asymptotics of (2.25) depends upon whether $1+2 \delta \, l(\alpha)$ 
exceeds the critical value 2, resulting in three possible limiting behaviors.

\medskip

\noindent
(a) When $2\delta\, l(\a)<1$, the second term in (2.25) converges to 0. 
Substituting in the limiting behavior (1.15) for $m_{|\a|}$,
we obtain
$$
C_{I,J}\Phi(\ax,z)\to M_{|\a|} c_{I,J}e^{-{1\over 2}\a\cdot 
X}e^{\delta\,\omega_2
(1-l(\a))}.
$$
The only non-zero contributions in the limit $\omega_2\rightarrow\infty$ arise
for simple roots $\alpha$,
$$
C_{I,J}\Phi(\ax,z)
\rightarrow
\cases{M_{|\a|} c_{I,J}e^{-\12\a\cdot X},& if $l(\a)=1$;\cr
$0$,& otherwise.\cr}
\eqno (2.26)
$$

\medskip

\noindent
(b) When $2\delta\, l(\a)>1$, only the second term in (2.25) survives and we
find
$$
C_{I,J}\Phi(\ax,z)
\to
-{M_{|\a|}}c_{I,J}e^{{1\over 2}\a\cdot X}
e^{\omega_2(\delta\, l(\a)+\delta-1)}Z^{-1}.
$$
Two cases arise : for $\delta < 1/h_\G$, the above quantity vanishes for
all roots $\alpha$. For $\delta= (l_0+1)^{-1} = 1/h_\G$, the
right hand side will vanish for all roots, except for the highest root $\alpha
_0$. Thus, it follows immediately that in this case
$$
C_{I,J}\Phi(\ax,z)
\rightarrow
\cases{-M_{|\a|} c_{I,J}Z^{-1} e^{{1\over 2}\a_0\cdot X} ,
& if $l(\a)=l(\alpha _0)$\cr
0, & otherwise.\cr}
\eqno (2.27)
$$

\medskip

\noindent
(c) When $2\delta\, l(\a)=1$, both terms in (2.25) have the same asymptotic
behavior as $\Re (\omega _2) \to \infty$, which is proportional to $\exp \{
\omega _2 (\delta - \half) \}$.  
For $\delta < \half$, or equivalently $l_0 >1$, this
factor, and thus $m_{|\a|}\Phi(\ax,z)$ tends to $0$. For $\delta = \half$, or
equivalently $l_0=1$ and $h_\G =2$, the simple Lie algebra must be $\G =
A_1=B_1=C_1$. The only positive root $\alpha$ (which is the highest root), 
yields
$$
C_{I,J}\Phi(\ax,z)
\to
{M_{|\a|}}
c_{I,J}(e^{-\half \alpha \cdot X}
-Z^{-1}e^{\half \a\cdot X}).
\eqno (2.28)
$$

\bigskip

\noindent
{\it Negative Roots}

When $l(\a)<0$, the second term in (2.25) is always negligible compared to the
first. The limit of $C_{I,J} \Phi(\ax,z)$ then rather depends on
whether $z-\ax$ tends to $-\infty$, $+\infty$, or remains finite.
This corresponds to the three cases $2\delta \, l(\a)>-1$,
$2\delta \, l(\a)<-1$, $2\delta \, l(\a)=-1$, which we examine in turn.

\medskip

\noindent
(a) When $2\delta \, l(\a)>-1$, the limit of $C_{I,J}\Phi(\ax,z)$ is given by
$$
C_{I,J}\Phi(\ax,z)
\to
-{M_{|\a|}}c_{I,J}
e^{\omega_2(\delta+\delta \, l(\a))}
e^{{1\over 2}\a\cdot X},
$$
which admits a non-vanishing limit only when $l(\a)=-1$:
$$
C_{I,J}\Phi(\ax,z)\rightarrow
\cases{-{M_{|\a|}}
c_{I,J}e^{{1\over 2}\a\cdot X}, & if $l(\a)=-1$\cr
0, & otherwise.\cr}
\eqno (2.29)
$$

\medskip

\noindent
(b) When $2\delta \, l(\a)<-1$, we have the following limiting behavior
$$
C_{I,J}\Phi(\ax,z)
={M_{|\a|}} Z
c_{I,J}e^{\omega_2(\delta- \delta \, l(\a)-1)}
e^{-{1\over 2}\a\cdot X},
$$
which admits a non-vanishing limit only when $\delta = 1/h_\G$ and
$l(\a)=-l_0$:
$$
C_{I,J}\Phi(\ax,z)\rightarrow
\cases{{M_{|\a|}} Z
c_{I,J}e^{-{1\over 2}\a_0\cdot X}, & if $l(\a)=-l_0$ and $\delta = 1 /h_\G$ \cr
& \cr
0, & otherwise.\cr}
\eqno (2.30)
$$

\medskip

\noindent
(c) When $2\delta \, l(\a)=-1$, $C_{I,J}\Phi(\ax,z)$ scales as $\exp \{ \omega
_2 (\delta - \half)\}$, which tends to 0 unless $\delta =\half$, $l_0=1$,
$h_\G=2$ and thus $\G = A_1, B_1, C_1$. In this case, the only root $\alpha$
yields
$$
C_{I,J} \Phi(\ax,z)
={M_{|\a|}} c_{I,J}
(Z e^{-{1\over 2}\a\cdot X} - e^ {\half \alpha \cdot X}).
\eqno (2.31)
$$

\medskip

In summary, for Lie algebras with $h_\G \geq 3$, we have found that
$$
C_{I,J}\Phi(\ax,z)\rightarrow
\cases
{\pm M_{|\a|} c_{I,J} e^{\mp\12\a\cdot X}, & if $l(\a)=\pm 1$;\cr
\mp M_{|\a|} c_{I,J} e^{\pm\12\a_0\cdot X}Z^{\mp 1}, & if $l(\a)=\pm l_0$ and
$\delta = 1 /h_\G$; \cr 
0 & otherwise.\cr}
\eqno(2.32)
$$
The case $h_\G=2$ for $\G=A_1,B_1,C_1$ may be read off from (2.28) and (2.31).

\medskip

We turn now to the limit of $C_{I,J}\Phi'(\ax,z)$. Replacing 
$C_{I,J}\Phi(u,z)$
by its approximation (2.24), we may write
$$
C_{I,J}\Phi'(u,z)
=C_{I,J}\Phi(u,z)\big[\half {\rm coth\,}{z\over 2}
+\partial_u\l\T^*(z-u|\tau)
-\half {\rm coth\,}{u\over 2}\big]
\eqno(2.33)
$$
Thus we need only determine
the limit of $\partial_u\l\T^*(z-u|\tau)$. It is
readily seen that 
$$
\partial_u\l\T^*(z-u|\tau)
\rightarrow
\cases{+\12, & $\qquad  \delta \, |l(\a)|<\12$\cr 
{3\over 2}, & $\qquad  \delta \, l(\a)>\12$\cr
-\12,        & $\qquad  \delta \, l(\a)<-\12$.\cr}
\eqno(2.34)
$$
Putting all together, we arrive at
$$
\eqalign{
\lim C_{I,J} & \Phi'(\ax,z)
=  
\half\epsilon_{\a}  {\rm lim} C_{IJ}\Phi(\ax,z), \cr
\epsilon_{\a} = & \left \{
\matrix{ 
+1 & l(\a)= + l_0 \ {\rm or}\ l(\a)=-1, \cr
-1 & l(\a)= - l_0 \ {\rm or}\ l(\a)=+1, \cr} \right\} .
\cr}
\eqno(2.35)
$$
While the derivation was carried out for $\delta \, |l(\alpha
)|\not= \half$ using (2.34), the final result (2.35) holds also for this case.

\medskip

It remains to discuss the limit of the operator
$D$ in (2.19). Its detailed structure was given in [1]. Here, the only 
information we shall need about it is that all contributions to
$D$ are of the form $m_{|\alpha|} \wp  (\ax)$ for some set
of roots $\alpha$. We note that in the expressions for
the entries of $D$ derived in [2], constants independent
of $\ax$ may be dropped, because the equations for the
entries of $D$ involve only differences in $\wp$.
Thus we may ignore constants such as ${\eta_1\over\pi i}$
in (2.3), just as in the case of limits of Hamiltonians.
Then the key observation is that the coefficients $C_{I,J}$ in the Lax pair
are proportional to a {\it single} power of the Calogero-Moser coupling
constants $m_{|\alpha|}$, while in the Hamiltonian the analogous
coefficients occur with the power 2. Now, we have already shown that
in the Hamiltonian each of these contributions admits a finite
limit with power 2. As $m_{|\alpha|} \to \infty$ in all cases, 
we see right away that $D \to 0$ in the limit.

\medskip

Combining the results summarized in (2.29) with (2.32), and 
using the above result that $ D \to 0$, we 
recover precisely the Lax pair with spectral parameter
for the affine Toda system when $\delta = 1/h_\G$, and the 
Lax pair for the ordinary Toda system when $\delta < 1/h_\G$.
The explicit forms may be derived by combining (2.19), (2.20), 
(2.21), (2.32) and (2.35) and we find
$$
\eqalign{
L_T & = \sum_{i=1}^nP_ih_i + \sum _{\alpha \in \R _* (\G)} 
M_{|\alpha|}
e^{-\half \alpha \cdot X} \bigl (E_\alpha - E_{-\alpha} \bigr ) 
+ M_{|\alpha _0|} 
e^{+\half \alpha_0 \cdot X} \bigl (- Z^{-1} E_{\alpha _0} 
       + Z   E_{-\alpha_0} \bigr )
\cr
M_T & = -\half  \sum _{\alpha \in \R _* (\G)} 
 M_{|\alpha|}
e^{-\half \alpha \cdot X} \bigl (E_\alpha + E_{-\alpha} \bigr ) 
 +
\half M_{|\alpha _0|} 
e^{+\half \alpha_0 \cdot X} \bigl ( Z^{-1} E_{\alpha _0} + Z E_{-\alpha_0}
\bigr ).
\cr}
\eqno (2.36)
$$
with the following conventions.
The summation is over the set $\R _* (\G)$ of simple roots of $\G$. When
$\delta = 1 /h_\G$, and $M_{|\alpha _0|}\not=0$, we have the affine Toda system
associated with the untwisted affine Lie algebra $\G ^{(1)}$, where $-\alpha 
_0$
plays the role of the extra affine root. When $0<\delta <1/h_\G$, and
$M_{|\alpha _0|}=0$, we have the ordinary Toda system associated with the
finite-dimensional Lie algebra $\G $. The matrices $E_\alpha$ are expressed in
terms of the constants $c_{I,J}$ of (2.21), and the generators $E_{IJ}$ of 
$GL(N,
\C)$, $I,J=1,\cdots ,N$ by
$$
E_\alpha = \sum _{I\not= J ;\alpha _{IJ} =\alpha}
c_{I,J} E_{IJ}.
\eqno (2.37)
$$
The Lax equation $\dot L_T = [L_T, M_T]$ is equivalent to the Hamilton-Jacobi
equations for the Toda Hamiltonian of (1.10).

\bigskip
\bigskip

\centerline{\bf III. TWISTED CALOGERO-MOSER AND
AFFINE TODA SYSTEMS}

\bigskip

We have established previously that the root system of each simple Lie algebra 
$\G$ defines an elliptic Calogero-Moser system, with Hamiltonian 
$H_{CM}$, given in (1.11). In the limit where the Calogero-Moser coupling $m$ 
tends to $\infty$, the system tends to an affine Toda system associated with 
the untwisted affine Lie algebra $\G ^{(1)}$. The coupling $m$ and the modulus 
$\tau$ of (1.1) then scale according to 
$$
m= M e^{ \omega _2 \delta}
\qquad \qquad 
\delta = 1/h_\G, 
\eqno(3.1)
$$
where $h_G$ is the Coxeter number of $\G$. 

\medskip

However, if $m$ is to correspond to the mass of a hypermultiplet in the adjoint 
representation for an $\N=2$ supersymmetric $\G$ gauge theory, then 
considerations based on the renormalization group behavior, on $R$-symmetry and 
on instanton calculus [16-19] require that the hypermultiplet decouple
rather according to (1.9), or equivalently, according to the following scaling 
rule
$$
m=M e^{ \omega _2 \delta ^\vee}
\qquad \qquad 
\delta ^\vee  = 1/ h ^\vee _\G,
\eqno(3.2)
$$
where $h_G^{\vee}$ is the {\it dual Coxeter number}. For simply-laced
algebras (i.e. when all roots have equal length, c.f. Table 2 in [1]), we 
have $h_\G = h ^\vee _\G$. However, for non-simply laced $\G$ (i.e. when $\G$ 
has roots of unequal length), we have instead $h^\vee _\G < h _\G$ and thus 
$\delta ^\vee > \delta$.
In this case the elliptic Calogero-Moser systems (1.11) do not scale to a 
finite limit. Thus, the untwisted Calogero-Moser systems are not expected to be 
the correct integrable systems associated with ${\cal N}=2$ supersymmetric 
Yang-Mills 
theories with adjoint hypermultiplet when the gauge group $\G$ is non-simply 
laced.

\medskip

This situation led us to introduce new, so-called {\it twisted} 
Calogero-Moser systems in [1], which are associated with non-simply laced 
$\G$, and whose Hamiltonians are given by (1.12).
We shall show in this section that these {\it twisted} Calogero-Moser
systems associated with non-simply laced Lie algebras $\G$ scale to a finite 
limit under (3.2). Furthermore, their limits are
affine Toda systems associated with the affine Lie algebras 
$(\G^{(1)})^{\vee}$,
that is, the dual of the untwisted affine Lie algebra $\G^{(1)}$.
 We begin by briefly reviewing the key features of our construction. For more 
details, see [1].
 
\medskip

$\bullet$ In the twisted Calogero-Moser Hamiltonians the short roots of 
$\G $ are twisted by replacing $\wp(\alpha\cdot x)$
with $\wp_\nu(\alpha\cdot x)$, and where $\nu$ equals the ratio of the ${\rm
length}^2$ of the long to the short roots. Clearly, we only need the
values $\nu  =1,2,3$. The functions $\wp_\nu(u)$ are defined in (1.13).

\medskip

$\bullet$ The scaling law for the dynamical variables $x$ (which was previously 
$x=X+2\omega_2\delta \rho^{\vee}$) is to be replaced
by $x= \xi (X+2\omega_2\delta^{\vee} \rho)$, where $\rho$ is now the Weyl 
vector of $\G$, $\dv$ is a scaling exponent and $\xi$ is a normalization 
dependent parameter. When all long roots $\alpha$ are normalized so that 
$\alpha ^2=2$, we have $\xi =1$ for all Lie algebras, as indicated in Theorem 
2. However, it is convenient to normalize the long roots $\alpha$ of $C_n$ to 
$\alpha ^2=4$. As a result, the normalization we shall use leads to $\xi=1$ for 
$\G = B_n, F_4, G_2$ and $\xi =\12$ for $\G=C_n$.

\medskip

Let $\R_s(\G)$ and $\R_l(\G)$ denote respectively the set of
short roots and the set of long roots of $\G$. Our choice of normalization
is given in Table 2 of [1]. Note that in all cases except $\G = C_n$,
the long roots $\alpha$ are normalized to have $\alpha ^2=2$,
while for $C_n$, they have $\alpha ^2 =4$. The twisted elliptic 
Calogero-Moser Hamiltonian associated to $\G$, and defined in (1.12) is then 
given by
$$
H_{TCM}=
\half p \cdot p
-\half\sum_{\a\in \R_s(\G )}m_s^2\wp_\nu (\a\cdot x)
-\half\sum_{\a\in \R_l(\G  )}m_l^2\wp(\a\cdot x)
\eqno(3.3)
$$
We shall determine the limit of $H_{TCM}$ under the scaling rule (3.2), but we
shall also allow for the scaling exponent $\dv < 1/h^\vee _\G$, for the sake of
completeness. 

\medskip

It is very useful to introduce the {\it dual level} function $l^\vee  
(\alpha)$,
defined by
$$
l^\vee (\alpha) \equiv \alpha \cdot \rho.
\eqno (3.4)
$$
This function is relevant here because the new scaling law for $x$ mentioned 
above naturally appeals to the dual level with $\alpha \cdot x \sim \alpha 
\cdot X + 2 \omega _2 \dv \ra.$ For a systematic exposition, see [1].
In terms of the decompositions of a root $\alpha$ and its coroot $\alpha ^\vee 
= 2\alpha /\alpha ^2$ onto simple roots and co-roots with integer 
coefficients $l_i$ and $l_i ^\vee$ (c.f. \S I), we have
$$
\alpha  = \sum _{i=1} ^n l_i \alpha _i, 
\qquad
\alpha ^\vee  = \sum _{i=1} ^n l _i ^\vee \alpha _i ^\vee,
\qquad
l_i ^\vee = { \alpha _i ^2 \over \alpha ^2} l_i ,
\qquad
l^{\vee}(\alpha)=\sum_{i=1}^n{\alpha_i^2\over 2}l_i.
\eqno (3.5)
$$
It is suggestive to consider $l^{\vee}$ as a function
of the {\it coroots} $\alpha^{\vee}$ rather than of the roots $\alpha$
$$
l^{\vee}(\alpha^{\vee})
={2\over\alpha^2}l^{\vee}(\alpha)
=\sum_{i=1}^nl_i^{\vee}.
$$
Then $l^{\vee}(\alpha^{\vee})$ satisfies the following properties
which are analogous to the properties of $l(\alpha)$
required in the evaluation of the limit of the untwisted Calogero-Moser system:

\medskip
\item{(1)} The minimal value $l^\vee (\alpha ^\vee)=1$ on positive coroots 
is attained
if and only if $\alpha$ is a simple root.  
\item{(2)} The maximal value of $l^\vee(\alpha^{\vee})$ on positive coroots is
$l^\vee (\alpha ^\vee ) = h_\G ^\vee -1= l_0 ^\vee$,
and it is attained if and only if $\alpha =
\alpha _0$ is the highest root of $\G$. 
\item{(3)} The highest root $\alpha _0$ is always
long, while its coroot $\alpha _0 ^\vee$ is always short.

\medskip

By reasoning that parallels the discussion in \S II.A,  one
argues that the convergence of the Hamiltonian under the scaling limit of 
(1.16) 
forces
$x$ to be shifted by a function linear in $\omega _2$ and the Weyl vector for
{\it maximally symmetric limits.} The additional constraints from requiring
that the $n\not=0$ terms converge may then be simply satisfied by requiring
that $0<\dv \leq \dv l^\vee (\alpha ^\vee) \leq 1-\dv$ for all roots $\alpha$
of $\G$. This condition is equivalent to $\dv \leq 1/h^\vee _\G$.
Assuming this condition, it is useful to recast (1) and (2) above
as

$$
\delta^{\vee}\leq \delta^{\vee}l^{\vee}(\alpha^{\vee})
\leq 1-\delta^{\vee},
\eqno (3.6)
$$ 
with equality on the left if and only if $\a$
is a simple root, and equality on the right if and only if
$\delta=1/h_{\G}^{\vee}$, and $\a$ is the longest root $\alpha_0$. 

\medskip

The limits of the Hamiltonian $H_{TCM}$ and of the Lax operators (to be
analyzed in \S IV) of the twisted Calogero-Moser systems will be taken 
according
to
$$
\eqalignno{
e^z             &= Z ^\xi e^{-\omega _2}\cr
m_{|\alpha|}    &= M_{|\alpha|} e^{\omega _2 \delta ^\vee} \cr
x               &= \xi (X+2\omega_2\dv \rho) \cr
\alpha \cdot x  &= \xi (\alpha \cdot X + 2 \omega _2 \dv l^\vee (\alpha))
&(3.7)\cr}
$$
where $Z,M_{|\alpha|}$ and $X$ are kept fixed, and the dual level $l^\vee 
(\alpha)$ of a root $\alpha$ was defined in (3.4). The factor $\xi$ is defined
by $\xi=1$ for $\G = B_n, F_4, G_2$ and $\xi=\half$ for $\G = C_n$. It is 
necessary
because only for $C_n$ is the normalization of the long roots $\alpha ^2=4$
instead of $2$. We shall discuss in detail only the case of positive roots. 

\medskip

In the subsections below, we shall establish that for the twisted
Calogero-Moser systems associated with any (non-simply laced) Lie algebra
$\G$, the potential terms in the Hamiltonian
$H_{TCM}$ converge to the following limits,
$$
m_{|\alpha|} ^2\wp _\nu (\ax)
\rightarrow
M_{|\alpha|} ^2
\cases{e^{\mp \a ^\vee \cdot X}, & if $l ^\vee(\alpha ^\vee)=\pm 1$;\cr
e^{\pm \a_0 ^\vee \cdot X},& if $l^\vee (\alpha  ^\vee)=\pm l_0 ^\vee$ and
$\dv = 1/h^\vee _\G$;\cr
0, & otherwise.\cr}
\eqno (3.8)
$$
This means that, in the limit (3.7), $H_{TCM}$
converges to the Hamiltonian of a Toda system associated to a Lie algebra for
which the simple roots are the coroots of $\G$, augmented by the negative
of the coroot $\alpha _0 ^\vee$ of the highest root $\alpha _0$ of $\G$. The
coroot $-\alpha _0 ^\vee$ plays the role of the affine root for the dual affine
Lie algebra $(\G ^{(1)})^\vee$. Thus, proving (3.8) will indeed prove that the
twisted Calogero-Moser Hamiltonian $H_{TCM}$ for the finite dimensional
Lie algebra $\G$ converges to the Toda Hamiltonian
$H_T$ for either the affine Lie algebra $(\G ^{(1)})^\vee$ when $\dv = 1/h_\G
^\vee$ or for the finite dimensional Lie algebra $\G ^\vee$ when $\dv < 
1/h_{\G}
^\vee$, establishing Theorem 2 (a) and (b) for the Hamiltonian.  

\medskip

Although the arguments are essentially the same for all non-simply laced simple
algebras, it is convenient to discuss separately the case of 
$\G = B_n,F_4$,
the case of $\G =G_2$, and the case of $\G =C_n$, since their values of $\nu$ 
and
their normalizations differ. 

\bigskip

\noindent {\bf (a) Twisted  Elliptic Calogero-Moser for $\G=B_n,
F_4$}

\medskip

These cases are characterized by the fact that the ratio of ${\rm length}^2$
of the roots is 2, and that the long roots $\alpha$ of $\G$ are normalized 
to $\alpha ^2 =2$, so that $\xi=1$ in (3.7). 

\medskip

We begin by analyzing the contributions of the (positive) long roots. 
Applying 
the asymptotics for $\wp(z)$ in terms of hyperbolic 
functions established in (2.3), we find as in (2.12)
$$
\eqalign{
m_l ^2 \wp(\ax)
\to
M_l ^2 \bigl [ &
\sum_{n\leq\dv \ra}
e^{-2 \omega _2(-n+\dv \ra) -\dv )-\a\cdot X}
\cr
+ &
\sum_{n>\dv \ra}
e^{-2 \omega _2(+n-\dv \ra -\dv) +\a\cdot X}\big].
\cr}
$$
Since the long roots are normalized to $\alpha^2=2$,
they satisfy $\alpha=\alpha^{\vee}$, and $l^{\vee}(\alpha)
=l^{\vee}(\alpha^{\vee})$. 
For reasons that will become completely clear when we
deal with the short roots, we prefer to recast all expressions
below in terms of coroots.
Now we are assuming that $\dv \leq 1/h ^\vee _\G$. Thus we have $\dv \leq
\dv l^{\vee}(\alpha^{\vee})=\dv\ra \leq 1-\dv$, and the above limit
further  reduces to
$$
m_l ^2 \wp(\ax)
\to
M_l ^2\big[
e^{-2 \omega _2 (\dv l^{\vee}(\alpha^{\vee}) - \dv ) -\a\cdot X}
+e^{-2 \omega _2 ( 1-\dv l^{\vee}(\alpha^{\vee}) -\dv) + \a\cdot X}\big]
$$
Clearly, the limit of the right hand side is then always finite, and non-zero
only when either $l^{\vee}(\alpha^{\vee})=1$ or 
$l^{\vee}(\alpha^{\vee})+1=h_\G^{\vee}$. This is the case exactly
when $\a$ is simple, or when $\a$ is the highest root 
$\alpha_0$ of $\G$. We thus recover (3.8) for long positive roots.

\medskip

We turn next to the short roots of $\G $. In view of (2.3), the function 
$\wp_2(u)$ can be expressed as (dropping irrelevant additive constants)
$$
\wp_2(u)=2\sum_{n=-\infty}^{\infty}
{1\over {\rm cosh}(2u-4n\o_2)-1}
\eqno(3.9)
$$
The condition $\dv \leq \dv l^\vee (\alpha ^\vee) \leq 1-\dv$,
on short roots, for which $\alpha ^\vee = 2 \alpha$, becomes $\dv \leq
2\dv l^\vee (\alpha ) \leq 1-\dv$.  The leading terms in
$m_s ^2\wp_2(\ax)$ are thus given by
$$
m_{|\alpha |} ^2 \wp_2(\ax)
\to 4 M_{|\alpha |} ^2 \bigl [
e^{-2 \omega _2 (2\dv \ra -\dv) - 2 \alpha \cdot X}
+
e^{-2 \omega _2 (2 - 2\dv \ra -\dv) + 2 \alpha \cdot X}
\bigr ]
\eqno(3.10)
$$
The first term has a non-zero limit if and only if $2\ra=1$, that is, if
$l^{\vee}(\alpha^{\vee}=1$ and
$\a$ is a simple short root. The exponent in the second term involves
$2-2\dv \ra -\dv =2-\dv l^{\vee}(\alpha^{\vee})-\dv\geq 1$ for all short roots 
of 
$\G$, so that the second
term always  tends to zero in the limit (3.7). Recasting the final expression 
in
terms of coroots, we see that the novel factors of 2 in (3.10) get nicely
absorbed into the definition of coroots $\alpha ^\vee = 2 \alpha$ of short 
roots
$ \alpha$. Putting all together, we find that for all roots of $\G$, we have
formula (3.8).

\medskip

{}From inspection of the limit of (3.9) when $\dv=0$, it follows immediately 
that
the limit gives then the trigonometric Calogero-Moser system for the dual 
finite
dimensional Lie algebra $\G ^\vee$, thus establishing (c) of Theorem 2, for the
Hamiltonian. 

\bigskip

\noindent{\bf (b) Twisted Elliptic Calogero-Moser for $\G = G_2$}

\medskip

The arguments for the long roots of $G_2$ are identical to those for the
long roots of the twisted $F_4$ and $C_n$ cases. Since
the long roots of $G_2$ have $\alpha ^2=2$
with our normalization, they equal their coroot, and the
limits may be expressed as in (3.8) as well.

\medskip

Next, we concentrate on the short roots of $G_2$, and make use of the following
expansion for $\wp_3(u)$, which is also an easy consequence of (2.3) (dropping
irrelevant additive constants)
$$
\wp_3(u)
=
{9\over 2}
\sum_{n=-\infty}^{\infty}
{1\over {\rm cosh}(3u-6n\o_2) - 1}
\eqno(3.11)
$$
For short roots $\alpha$ of $G_2$, we have now $\alpha ^\vee = 3
\alpha$, and the condition $\dv \leq \dv l^\vee (\alpha ^\vee) \leq 1 - \dv$
becomes $\dv \leq 3\dv l^\vee (\alpha ) \leq 1 - \dv$. The leading terms in
$m_s ^2\wp_3(\ax)$ can be written as
$$
m_s ^2\wp_3(\ax)
\to 
{9\over 4} M _s ^2 
\big [
e^{ -2 \omega _2 (3 \dv \ra - \dv)-3\a\cdot X}
+e^{ - 2 \omega _2 (3-3 \dv \ra - \dv) + 3\a\cdot X}\big ].
\eqno(3.12)
$$
These lead to non-vanishing limits only when $3 \ra = l^\vee
(\alpha ^\vee)=1$, which means that $\a$ is a simple short root.
Since $3-3\dv\ra-\dv=3-\dv l^{\vee}(\alpha^{\vee})-\dv\geq 2$, the second term 
in 
(3.12) always converges to 0.
Thus only simple short roots survive from the sum over all short
roots in the Hamiltonian, and we again recover the result of (3.8).

\medskip

{}From the asymptotics of (3.11) in the limit (3.6), it is clear that this 
limit 
gives the trigonometric Calogero-Moser system for $G_2 ^\vee=G_2$, when 
$\dv=0$,
establishing (c) of Theorem 2, for the Hamiltonian.

\bigskip

\noindent{\bf (c) Twisted Elliptic Calogero-Moser for $\G = C_n$}

\medskip

The only difference separating this case from the earlier
ones is a difference of convention. Since our choice of longest roots
for $C_n$ is $2e_i$, $1\leq i\leq n$, the coroot of a long root $\alpha$ obeys
$\alpha ^\vee = \12 \alpha$, while the coroot of a short root $\alpha$
obeys $\alpha ^\vee = \alpha$.
Thus with the Weyl vector $\rho$ still defined by the same formula
$\rho=\sum_{i=1}^n\lambda_i$,
where $\lambda_i$ are the fundamental weights, the choice of 
scaling of the dynamical variables $x$ is now as in (3.7) with $\xi=\12$,
$$
x= \half (X+ 2\o_2\dv \ra).
\eqno(3.13)
$$

\medskip

{}First, we consider the contributions of the long roots $\alpha$ of $C_n$, for
which $\alpha ^\vee = \12 \alpha$. The condition $\dv \leq \dv l^\vee (\alpha
^\vee) \leq 1 - \dv$ then becomes $\dv \leq \half \dv l^\vee (\alpha
^\vee) \leq 1 - \dv$. The expansion of
$m_l^2\wp(\ax)$ is given by
$$
\eqalign{
m_l^2\wp(\ax)
\to
M_l ^2\big[ &
\sum_{\12\dv\ra>n}
e^{-2\omega _2 (-\dv+\12\dv\ra-n)-\12 \a\cdot X}
\cr
+ &
\sum_{n>\12\dv\ra}
e^{-2 \omega _2 (-\dv-\12\dv\ra+n) + \12 \alpha \cdot X}
\big ], \cr}
\eqno (3.14)
$$
and reduces to the following asymptotics
$$
\eqalign{
m_l^2\wp(\a\cdot x)
\to &
M^2 \bigl [
e^{-2 \omega _2(\dv\12\ra-\dv)-\12 \a\cdot X}
+e^{-2 \omega _2 (1-\12\dv\ra-\dv) +\12 \a\cdot X}
\bigr ], \cr
\to &
M^2 \bigl [
e^{-2 \omega _2\dv(l^\vee (\alpha ^\vee) -1)- \a ^\vee \cdot X}
+e^{-2 \omega _2 (1-\dv l^\vee (\alpha ^\vee) -\dv) + \a ^\vee \cdot X}
\bigr ].
\cr}
\eqno(3.15)
$$
Here, we have re-expressed the right hand side of the last line in terms 
of coroots.
These two terms produce a non-vanishing limit respectively when
$\a ^\vee $ is a simple coroot with $l ^\vee (\alpha ^\vee)=1$, so that $\alpha 
$
is a simple root, and when $\a^\vee $ is the highest coroot $\a_0^\vee$,
characterized by $l^\vee (\alpha ^\vee _0) = l_0 ^\vee =h^{\vee} _\G -1$. We 
thus
recover, for long roots of $C_n$, the result announced in (3.8).

\medskip

Next, we consider the contribution from short roots. Using (3.9) and (3.13), 
the
combination $m_s^2\wp_2(\ax)$ is given by 
$$
m_s^2\wp_2(\ax)
=
\sum_{n=-\infty}^{\infty}
{ 2 M_s ^2e^{2 \omega _2 \dv}\over  \cosh (\a\cdot X+2 \o_2\dv \ra-4n\o_2)-1}
\eqno(3.16)
$$
and the leading terms are
$$
m_s^2\wp_2(\ax)
\to
4 M_s ^2\big [
e^{-2 \omega _2 (-\dv+\dv\ra) -\a\cdot X}
+e^{- 2 \omega _2 (-\dv-\dv\ra+2) + \a\cdot X}
\big ]
\eqno (3.17)
$$
The first term on the right hand side produces a non-vanishing limit
exactly when $\ra=1$, which means in this case that $\a ^\vee = \a$
is a simple (co)root. The second term does not contribute in the limit,
since $2-\dv-\dv\ra\geq 1$. Expressing the full answer in terms of coroots, we
again recover (3.8).

\medskip

{}From the asymptotics of (3.11), it is clear that the limit (3.6) gives the
trigonometric Calogero-Moser system for $C_n ^\vee=B_n$, when $\dv=0$, thus
establishing (c) of Theorem 2, for the Hamiltonian. 

\bigskip
\bigskip

\centerline{\bf IV. LIMITS OF LAX PAIRS }

\bigskip

To complete the proof of Theorem 2, we shall establish in this section the 
limits
of the Lax pairs according to the scaling limit (3.7).
The Lax pairs for the twisted Calogero-Moser systems were constructed
explicitly in [1], and are of the form (2.19), but with a more general form
for $X$ and $Y$
$$
\eqalign{
X= & \sum _{I,J=1;I\not= J} ^N C_{I,J}  
                             \Phi _{IJ} (\alpha _{IJ} \cdot x, z) E_{IJ} \cr
Y= & \sum _{I,J=1;I\not= J} ^N C_{I,J}  
                             \Phi _{IJ} ' (\alpha _{IJ} \cdot x, z) E_{IJ}. 
\cr}
\eqno (4.1)
$$
All other notations and conventions are as in \S IIB.
Explicit expressions for the constants $C_{I,J}$ and for the elliptic functions
$\Phi _{IJ}$ were constructed in [1]. The data needed about these for the
limits will be given in the subsections below. The main complication of the
twisted cases is that there are now several different elliptic  functions $\Phi
_{IJ}$, whose limits will have to be studied.

\medskip

One general result is worth deriving right away. Just as in the case of the
untwisted Calogero-Moser systems, the matrix $D$ entering the Lax operator
$M$ in (2.19) is a sum of terms  proportional to
$m_{|\alpha |} \wp (\alpha \cdot x)$. This combination is similar to the terms
that enter the Calogero-Moser Hamiltonians, except that the power of
$m_{|\alpha|}$ is 1 instead of 2. As $m_{|\alpha|} \to \infty$ in (3.7), it
immediately follows that
$$
D \to 0
\qquad \qquad
\Delta, \ d_j \to 0, \ j=1,\cdots ,N,
\eqno (4.2)
$$ 
so that the Lax operator $M$ reduces to $Y$. Henceforth, we shall restrict to
the study of the $X$ and $Y$ parts of the Calogero-Moser Lax operators. 

\medskip

In the subsections below, we shall establish the following limits of the
entries $X$ and $Y$ of the Lax operators, as the constants $C_{I,J}$
scale with $\omega _2$ according to their expressions derived in [1] in terms
of $m_{|\alpha|}$,
$$
C_{I,J} = M_{|\alpha|}e ^{\dv \omega _2} c_{I,J}
\eqno (4.3)
$$
(Recall that $\alpha=\lambda_I-\lambda_J$).
For the Lie algebras $B_n$ and $F_4$, we shall show that the 
entries of $X$ satisfy
$$
C_{I,J}\Phi _{I,J}(\ax,z)\rightarrow
\cases {
\pm \kappa_{\G}M_{|\a|}c_{I,J} ~ e^{\mp\half\a ^\vee \cdot X}, 
& if $l ^\vee (\a ^\vee )=\pm 1$;\cr
\mp \kappa_{\G}M_{|\a|}c_{I,J} ~ e^{\pm\half\a_0 ^\vee \cdot X}Z^{\mp 1}, 
& if $l^\vee (\a ^\vee)=\pm l_0 ^\vee$ and $\dv = 1 /h_\G ^\vee$; \cr 
0 & otherwise,\cr}
\eqno(4.4a)
$$
where $\kappa_{\G}$ are constants depending on the algebra $\G$,
with $\kappa_{B_n}=1$ and $\kappa_{F_4}=2$. In the cases
of $B_n$ and $F_4$, the entries of the matrix $Y$
scale in analogy with the untwisted case
$$
C_{I,J}\Phi _{I,J} ' (\ax,z)\rightarrow
\cases{
\,-{1\over 2}\kappa_{\G}M_{|\a|}c_{I,J} ~ e^{\mp\half\a ^\vee \cdot X}, 
& if $l ^\vee (\a ^\vee )=\pm 1$;\cr
\,-{1\over 2}\kappa_{\G}M_{|\a|}c_{I,J} ~ e^{\pm\half\a_0 ^\vee \cdot X}Z^{\mp 
1}, 
& if $l^\vee (\a ^\vee)=\pm l_0 ^\vee$ and $\dv = 1 /h_\G ^\vee$; \cr 
0 & otherwise.\cr}
\eqno(4.5a)
$$
The case of $C_n$ differs from the other cases only in minor details.
More precisely, the matrix $X$ scales in this case according to
$$
C_{I,J}\Phi _{I,J}(\ax,z)\rightarrow
\cases {
\pm 2 M_{|\a|}c_{I,J} ~ e^{\mp\half\a ^\vee \cdot X}, 
& if $l ^\vee (\a ^\vee )=\pm 1$;\cr
\mp 2 M_{|\a|}c_{I,J} ~ e^{\pm\half\a_0 ^\vee \cdot X}Z^{-\half\mp\half}, 
& if $l^\vee (\a ^\vee)= \pm l_0 ^\vee$, $\dv = 1 /h_\G ^\vee$,
$I<J$; \cr
\mp 2 M_{|\a|}c_{I,J} ~ e^{\pm\half\a_0 ^\vee \cdot X}Z^{\half\mp\half}, 
& if $l^\vee (\a ^\vee)= \pm l_0 ^\vee$, $\dv = 1 /h_\G ^\vee$,
$J<I$;\cr
0 & otherwise,\cr}
\eqno(4.4b)
$$
while the matrix $Y$ scales as
$$
C_{I,J}\Phi' _{I,J}(\ax,z)\rightarrow
\cases {
-2 M_{|\a|}c_{I,J} ~ e^{\mp\half\a ^\vee \cdot X}, 
& if $l ^\vee (\a ^\vee )=\pm 1$;\cr
-2 M_{|\a|}c_{I,J} ~ e^{\pm\half\a_0 ^\vee \cdot X}Z^{-\half\mp\half}, 
& if $l^\vee (\a ^\vee)= \pm l_0 ^\vee$, $\dv = 1 /h_\G ^\vee$,
$I<J$; \cr
-2 M_{|\a|}c_{I,J} ~ e^{\pm\half\a_0 ^\vee \cdot X}Z^{\half\mp\half}, 
& if $l^\vee (\a ^\vee)= \pm l_0 ^\vee$, $\dv = 1 /h_\G ^\vee$,
$J<I$; \cr
0 & otherwise.\cr}
\eqno(4.5b)
$$

\medskip

The full expression for the Lax operators may be worked out, just as we did in
(2.36) for the untwisted cases. In fact, the result is completely analogous,
except that the summation is over the set $\R _* (\G^\vee)$ of simple coroots 
of
$\G$ and over the coroot $\alpha _0 ^\vee$ instead of over the root $\alpha_0$.
When $\dv = 1 /h_\G ^\vee $, and $M_{|\alpha _0 ^\vee |}\not=0$, we have the
affine Toda system associated with the dual affine Lie algebra $(\G^{(1)})^\vee
$, where $-\alpha _0 ^\vee $ plays the role of the extra affine root. When
$0<\dv <1/h_\G ^\vee$, and $M_{|\alpha _0 ^\vee |}=0$, we have the ordinary 
Toda
system associated with the finite-dimensional Lie algebra $\G ^\vee $. The
matrices $E_{\alpha ^\vee} $ are expressed in terms of the constants $c_{I,J}$ 
of
(4.1) and (4.3) and the generators $E_{IJ}$ of $GL(N, \C)$, $I,J=1,\cdots ,N$
just as in (2.37).

\medskip

In order to derive (4.4) and (4.5), it is convenient to proceed separately for
each of the non-simply laced finite-dimensional Lie algebras $\G = B_n, C_n,
F_4$ and $G_2$, since the structure of the Lax operators of the associated
Calogero-Moser system is quite different in each case.

\bigskip
 
\noindent
{\bf (a) The Limit of the Twisted $B_n$ Calogero-Moser Lax Pair}

\medskip

This case is relatively the simplest among
the twisted Calogero-Moser cases, as its only new feature
is the appearance of a new function $\Lambda(2u,z)$.
The twisted Calogero-Moser Hamiltonian for $B_n $ admits a Lax pair
of dimension $N=2n$, with spectral parameter $z$ and two independent couplings
$m_s$  and $m_l$, given by (2.19), (4.1) and
$$
\eqalignno{
\Phi _{IJ}  (x,z) = & \left \{ 
\matrix{\Phi (x,z) & I-J \not= 0,\pm n \cr
         \Lambda (x,z) & I-J = \pm n \cr } \right . 
& (4.6a) \cr
C _{I,J}   = & \left \{ 
\matrix{m_l & I-J \not= 0,\pm n \cr
        m_s & I-J = \pm n \cr }  \right .
& (4.6b) \cr
}
$$
The function $\Lambda (2u,z)$ is defined by
$$
\Lambda(2u,z)={\Phi(u,z)\Phi(u+\o_1,z)\over\Phi(\o_1,z)}.
\eqno(4.7)
$$
and was studied in detail in Appendix \S B of [1].
We observe that the prescription (4.6) implies in particular
that the long roots $\a=\pm e_i\pm e_j$ of $B_n$
occur only in entries of the form $\Phi_{IJ}(\ax)=\Phi(\ax)$,
while the short roots $\a=e_i$ of $B_n$ emerge from
$I-J=\pm n$, and appear only in entries of the form $\Phi _{IJ} (\ax) = \Lambda
(2\ax)$. With the help of this observation and of the limits already evaluated 
in
\S IIB of this paper, it is easy to determine the limits of the Lax pair. 

\medskip

{}For the long roots, we use the asymptotics of $\Phi(u,z)$ in (2.25). 
Since $u=2\o_2\dv\ra+\a\cdot X$, we need to consider
three cases, according to whether $u-\omega _2$ tends to $-\infty,
0$ or
$+\infty$ respectively.

\medskip

\noindent
(a) $2\dv\ra<1$. In this case, as $C_{I,J}$ scales according
to (4.3), the right hand side of (2.25) reduces to the first term and we have
$$
C_{IJ}\Phi(\ax,z) \to
M_{|\alpha|}c_{IJ} e^{\omega _2 \dv (1-l^\vee (\alpha)) -\12\a\cdot X}.
\eqno(4.8)
$$
The limit (4.8) is non-zero exactly when $\ra = l^\vee (\alpha ^\vee)=1$, which
means that $\a$ is a simple long root of $B_n$.

\medskip

\noindent
(b) $2\dv\ra=1$. In this case both terms on the r.h.s. of (2.25) contribute
in the limit and we have 
$$
C_{IJ}\Phi(\ax,z)
\to 
M_{|\alpha|}c_{IJ} e^{\omega _2 (\dv - \12) - \12 \a\cdot X}(1-e^{\a\cdot 
X}Z^{-1})
\eqno (4.9)
$$
A non-zero limit requires that $\dv=\12$ which cannot happen for $B_n$, since
$\dv=1/( 2n-1)$.

\medskip

\noindent
(c) $2\dv\ra>1$. In this case, the second term on the r.h.s. in (2.25)
dominates the asymptotics of $C_{IJ}\Phi(\ax,z)$, producing
$$
C_{IJ}\Phi(\ax,z)
\to
-M_{|\alpha|}c_{IJ} e^{\omega _2 (-1+\dv+\dv \ra)}e^{\12\a\cdot X} Z^{-1}.
\eqno(4.10)
$$
A non-vanishing limit arises when $\dv + \dv\ra=1$, which means that
$\a = \a _0$ is the highest root (which is a long root). 

\medskip

Now, in all three cases above, we are dealing with the long roots $\alpha$ of
$B_n$, which are normalized so that $\alpha ^2=2$ and thus $\alpha ^\vee =
\alpha$. Recasting the results of (4.8), (4.9) and (4.10) in terms of coroots,
we readily recover the result of (4.4) for long roots of $B_n$.

\medskip

We turn next to the short roots of $B_n$, which are of the form $\a=e_i$.
The asymptotics of $\Phi(\o_1,z)$ follows directly from (2.24) and we have
$ \Phi(\o_1,z)\to \12 i$. Combining this result with the exressions for the
asymptotics of $\Phi$ in (2.25) and the definition of $\Lambda (2u,z)$ in 
(4.7),
we find
$$
\Lambda (2u,z) \to \left \{
\matrix{
+2e^{- u} (1 - Z^{-2} e ^{2u- 2\omega _2} ) & \qquad \Re (u) \to +\infty \cr
&\cr
-2e^{+ u} (1 - Z^2 e^{-2u -2\omega _2} )      & \qquad \Re (u) \to - \infty 
\cr}
\right .
\eqno (4.11)
$$
We still have the scaling law $u=\ax=\a\cdot X+2\o_2\dv\ra$ of (3.7) and,
assuming that $\a$ is a positive root, the asymptotics produces three cases
according to whether $u-\omega _2$ tends to $-\infty,0$ or $+\infty$
respectively.

\medskip

\noindent
(a) $2\dv\ra < 1$. In this case, the second term on the right hand side of
(4.11) converges to 0, and we are left with the contribution of only the first
term,
$$
C_{IJ}\Lambda(2\ax,z)
\to
2 M_{|\alpha|}c_{IJ} e^{\omega _2 (\dv - 2\dv\ra)} e^{-\a\cdot X}.
\eqno(4.12)
$$
This has a non-zero limit exactly when $2\ra=1$, that is, when $\a$
is a simple short root.

\medskip

\noindent
(b) $2\dv\ra=1$. As before, it is easily seen that although
both terms in (4.11) contribute to the limit, a non-vanishing limit arises only
when $\dv=1$. This can only occur
for the special case $B_1$.

\medskip

\noindent
(c) $2\dv\ra>1$. In this case, the second term in (4.11) dominates the
asymptotics and we have
$$
C_{IJ}\Lambda(2\ax)
\to
- 2M_{|\alpha|}c_{IJ}e^{\omega _2 ( -2 + \dv + 4\dv l^\vee (\alpha))}e^{\a\cdot 
X}Z^{-2}
\eqno(4.13)
$$
This always tends to 0, since $-2+\dv +4 \dv l^\vee (\a ^\vee) \geq \dv>0$.

\medskip

The coroots of short roots $\alpha$ of $B_n$ obey $\alpha ^\vee = 2 \alpha$.
Recasting the results obtained in (4.12) and (4.13) in terms of coroots, we
readily recover (4.4). Finally, (4.5) in the case of $B_n$
is derived using the derivative
expressions of (2.35) for $\Phi$ and the analogous expression for $\Lambda$,
or simpler still, of the asymptotic expansions (2.25) and (4.11).

\bigskip

\noindent
{\bf (b) The Limit of the Twisted $C_n$ Calogero-Moser Lax Pair}

\medskip

The twisted Calogero-Moser Hamiltonian for $C_n $ admits a Lax pair
of dimension $N=2n+2$, with spectral parameter and one independent couplings
$m$ given by (2.19), (4.1) and 
$$
\eqalignno{
\Phi _{IJ}  (\a _{IJ} \cdot x,z) = & 
\Phi _2 (\a _{IJ} \cdot x +\omega _{IJ},z) 
& (4.14a) \cr
C _{I,J}   = & \left \{ 
\matrix{m & I,J = 1,\cdots ,2n; I-J \not= \pm n \cr
        \sqrt 2 m  & I=1,\cdots ,2n; J=2n+1,2n+2; 
I\leftrightarrow J \cr
        2m& I=2n+1, J=2n+2; I \leftrightarrow J\cr }  \right .
& (4.14b) \cr}
$$
The constants $\omega _{IJ}$ obey cocycle conditions, and are defined only up
to shifts $\zeta _I$ resulting from shifts in the vector $x$. Both are given by
$$
\eqalign{
\omega _{JI} = & - \omega _{IJ} \cr
\omega _{IJ} + \omega _{JK} & + \omega _{KI}=  0 \cr
\omega _{IJ} \to & \omega _{IJ} + \zeta _I - \zeta _J.
\cr}
\eqno (4.15)
$$
The fact that the Lax equations for this Lax pair must reproduce the
twisted Calogero-Moser Hamilton-Jacobi equations requires that $\omega _{IJ}$
take values amongst the half periods $- \omega _2,0,+\omega _2$, up to shifts
$\zeta _I$. A convenient solution is given by
$$
\omega _{IJ} = 
\left \{ 
\matrix{0 & I\not=J =1, \cdots , 2n+1 \cr
        +\omega _2 & I=1,\cdots ,2n; J=2n+2 \cr
        -\omega _2 & J=1,\cdots ,2n; I=2n+2. \cr } \right . 
\eqno (4.16)
$$
Special care is needed in properly defining the normalizations of the roots
$\alpha _{IJ}$. We have 
$$
\eqalign{
\alpha _{IJ} &= \lambda _I - \lambda _J 
\quad \qquad I,J =1,\cdots, 2n+2 \cr
\lambda _i & = -\lambda _{n+i} = e_i, 
\quad i=1,\cdots ,n, 
\qquad
\lambda _{2n+1}  = \lambda _{2n+2} =0. \cr
}
\eqno (4.17)
$$
Thus, for $I,J =1,\cdots ,n$, the entries $\alpha _{IJ}$ yield the short roots
of $C_n$, while when either $I$ or $J$, but not both, equals $2n+1$ or
$2n+2$, the entries $\alpha _{IJ}$ yield {\it half} of the long roots of $C_n$.
The function $\Phi _2$ is the function $\Lambda $ of
(4.7), but for double the argument, defined by
$$
\Phi _2(u,z)=\Lambda (2u,z).
\eqno(4.18)
$$
The limit is taken according to (3.7).
$$
\eqalign{
x&  =\half ( X + 2 \omega _2\, \dv \, \rho), \cr
e^z & = Z ^\12 e^{-\omega _2} \cr}
\eqno (4.19)
$$
where $\rho$ is the Weyl vector. Notice the extra factors of $\12$ related to
the non-canonical normalization of the long roots of $C_n$.
The asymptotics of $\Phi _2$ follows directly from those of $\Lambda$ in 
(4.11),
but this time for the scaling limit of (4.19),
$$
\Phi _2 (u,z) \to \left \{
\matrix{
+2e^{- u} (1 - Z^{-1} e ^{2u- 2\omega _2} ) & \qquad \Re (u) \to +\infty \cr
&\cr
-2e^{+ u} (1 - Z e^{-2u -2\omega _2} )      & \qquad \Re (u) \to - \infty. \cr}
\right .
\eqno (4.20)
$$

\medskip

We shall assume in the subsequent discussion that the roots $\alpha$ are
positive. The case of negative roots can be treated by similar arguments.
We consider first the entries related to the positive short roots
of $C_n$, given by $\a=\alpha _{IJ}=\pm e_i\pm e_j$, $1\leq i\leq n$. These 
arise
only when both indices $I,J$ satisfy $1\leq I,J\leq n$, in which
case $\o_{IJ}=0$, and $\Phi_{IJ}(\ax,z)=\Phi_2(\ax,z)$. Using the
asymptotics of (4.20), we find
$$
C_{I,J} \Phi _2 (\alpha \cdot x,z) 
\to 
2 M_{|\alpha|}c_{I,J} e^{-\omega _2 (\dv l^\vee (\alpha) - \dv) -\half \alpha 
\cdot X}
(1 - Z^{-1} e^{\alpha \cdot X + 2 \omega _2 (\dv l^\vee (\alpha ) -1)})
\eqno (4.21)
$$ 
For short roots $\alpha$ of $C_n$, we have $\alpha ^\vee = \alpha$, and thus 
$\dv \leq \dv l^\vee (\alpha ) \leq 1-\dv$. As a result, the factor in
parentheses in (4.21) converges to 1. The remaining factors tend to zero unless
$\dv l^\vee (\alpha)=1$, i.e. $\alpha$ is a simple short root. Recasting the
result in terms of coroots, we recover (4.4) for the short roots of $C_n$.
Similarly, we find 
$$
C_{I,J}\Phi_{I,J}'(\alpha\cdot x,z)
\to
-2 M_{|\alpha|} c_{IJ}e^{-\half\alpha^{\vee}\cdot X},
$$
as written earlier in (4.5).

\medskip

Next we consider the entries related to the positive long roots $2e_i$ of 
$C_n$.
These arise either under the form $\Phi_2(\half\ax,z)$ (when either $I$ or $J$ 
is
between $1$ and $n$, and the other index is $n+1$), or under the form
$\Phi_2(\half\ax\pm\o_2,z)$ (when one of the indices $I$ or $J$ is between $1$ 
and
$n$, and the other index is $n+2$).

\medskip

In the first case, we have $\Phi_2(u,z)$, with $u$ given by
$$
u= \half \ax = {1 \over 4} \a \cdot X + \half \omega _2 \dv l^\vee (\alpha).
\eqno (4.22)
$$
As $\omega _2 \to \infty$, $u$ satisfies $u \to +\infty$ and $u-\omega _2 \to
-\infty$, since $\dv l^\vee (\a) < 2$, so that only the first term on the 
r.h.s.
in (4.20) remains in the limit, and we obtain 
$$
\eqalign{
C_{IJ}\Phi_2 (\half\ax,z)
\to &
2 M_{|\alpha|}c_{IJ} e^{\omega _2 (\dv-\half \dv l^\vee (\a))}e^{-\quarter 
\a\cdot 
X}
\cr
\to &
2 M_{|\alpha|}c_{IJ} e^{\omega _2 (\dv- \dv l^\vee (\a ^\vee))}e^{-\half \a 
^\vee 
\cdot X}.
\cr}
\eqno (4.23)
$$
Here, we have re-expressed the limit in terms of coroots on the second line.
The limit is non-zero if and only if $\a$ is a simple long root
for $C_n$, and we thus recover the result of (4.4). The limit for
$C_{I,J}\Phi_{I.J}'(\half\ax,z)$ in (4.5) follows then easily
from the asymptotics $\Phi(u,z)\sim 2 e^{-u}$ which apply in this case.

\medskip

Next we consider the case $\Phi_2(u,z)$ with $u=\half \ax +\omega _2$. As 
$\omega
_2 \to \infty$, we now have $u \to +\infty$, but $u-\omega _2 \to +\infty$ as
well. Thus, the limit is dominated by the second term on the r.h.s. of (4.20),
and we find 
$$
\eqalign{
C_{IJ}\Phi_2(\half\ax +\omega _2,z)
\to &
-2M_{|\alpha|}c_{IJ} 
e^{\omega _2 (\dv + \12 \dv l^\vee (\alpha) -1) + \quarter \a \cdot X}Z^{-1}
\cr
\to &
-2M_{|\alpha|}c_{IJ} 
e^{\omega _2 (\dv +  \dv l^\vee (\a ^\vee) -1) + \12 \a ^\vee \cdot X}Z^{-1}
\cr}
\eqno(4.24)
$$  
This has a non-vanishing limit only when $\dv+\dv l^\vee (\a ^\vee )-1=0$, 
i.e.,
when $\a$ is the highest root. In this case, the longest root is $2e_1$, and it
does indeed occur amongst these roots $\alpha$. Again, we recover the
results of (4.4), and just as easily, of (4.5) for
$C_{I,J}\Phi_2(\half\ax+\omega_2,z)$.

\medskip

{}Finally, to evaluate the limit of $C_{IJ}\Phi_2(\12\ax-\o_2,z)$,
it is easiest to make use of the monodromy properties of $\Phi_2(u,z)$
$$
\Phi_2(u-\o_2,z)=\Phi_2(u+\o_2,z)e^{-4(\o_2\zeta(z)-\eta_2z)}
$$
This relation becomes particularly simple when we set
$\o_1=-i\pi$, take the limit $\o_2\rightarrow\infty$, and use the
relation $\eta_1\o_2-\eta_2\o_1=\half i\pi$
$$
\Phi_2(u-\o_2,z)=\Phi_2(u+\o_2,z)Z
\eqno(4.25)
$$
Substituting in the limits found for $\Phi_1(\12 \ax+\o_2,z)$ in (4.24),
we conclude that the only non-vanishing limit occurs again
only for the highest root $2e_1$, with the values indicated in
(4.4). The derivative terms can be evaluated in the same way,
leading to (4.5), and our treatment of the $C_n$ case is complete.
\bigskip

\noindent
{\bf (c) The Limit of the Twisted $F_4$ Calogero-Moser Lax Pair}

\medskip

The twisted Calogero-Moser Hamiltonian for $F_4 $ admits a Lax pair
of  dimension $N=24$, with spectral parameter $z$ and two independent couplings
$m_s$  and $m_l$, whose form is given by  
$$
\eqalignno{
\Phi _{\lambda \mu}  (x,z) = & \left \{ 
\matrix{\Phi  (x,z) & \lambda \cdot \mu =0  \cr
        \Phi _1 (x,z) & \lambda \cdot \mu = \half \cr
        \Lambda (x,z) & \lambda \cdot \mu =-1 \cr } \right . 
& (4.26a) \cr
C _{\lambda , \mu}   = & \left \{ 
\matrix{m_l & \lambda \cdot \mu =0 \cr
        {1 \over \sqrt 2} m_s & \lambda \cdot \mu =\half \cr
        0   & \lambda \cdot \mu =-\half \cr
        \sqrt 2 m_s & \lambda \cdot \mu = -1 \cr }  \right .
& (4.26b) \cr
}
$$
Here, the entries are labeled by the 24 {\it 
non-zero} weights $\lambda$ of the {\bf 26} of $F_4$, which are also the 24 
short roots of $F_4$. The functions $\Lambda $ and $\Phi _1 $ are defined
respectively by (4.7) and
$$
\eqalign{\Phi_1(u,z)
=&
\Phi  (u,z)+f(z)\Phi  (u+\o_1,z)\cr
f(z)=&
-e^{\pi i \zeta(z)+\eta_1 z},
\cr}
\eqno (4.27)
$$
so that we have simple monodromy with period $\omega _1$, given by
$\Phi _1 (u +\omega _1,z) = f(z)  ^{-1} \Phi _1 (u,z)$.

\medskip

The above classification of $\Phi_{\lambda\mu}(u,z)$ depending on the values
of $\lambda\cdot\mu$ leads to the three possible ways in which roots
of $F_4$ can arise in the Lax pair:
when $\lambda\cdot\mu=0$, $\lambda - \mu$ is a long root, of the form
$\a=\pm e_i\pm e_j$; 
when $\lambda\cdot\mu=\12$, $\lambda -\mu$ is a short root; 
when $\lambda\cdot\mu=-1$,
$\lambda-\mu=2\alpha$, where $\alpha$ is again any of the short roots.
We also evaluate the limits separately in the three cases.

\medskip

{}For the long roots $\lambda\cdot\mu=0$, the discussion
is identical to that of the case $B_n$ in \S IV (a). We conclude that
the limit of $C_{\lambda\mu}\Phi_{\lambda\mu}(\ax)$ is non-zero
only when $\a$ is either a simple (long) root or the highest root. 
Long roots $\alpha $ of $F_4$ satisfy $\a ^\vee = \a$, so that
$$
C_{\lambda\mu}\Phi_{\lambda\mu}(\ax,z)
\rightarrow
\cases{
+2M_{|\alpha|}c_{\lambda\mu}e^{-\12\a ^\vee \cdot X}, & if $l^\vee (\a 
^\vee)=1$;\cr
-2M_{|\alpha|}c_{\lambda\mu} e^{\12\a _0 ^\vee \cdot X} Z^{-1}, & if $l^\vee
(\a ^\vee) +1=h^{\vee} _\G$,}
\eqno (4.28)
$$
reproducing (4.4) for long roots.

\medskip

Next, we evaluate the limit of $C_{\lambda\mu}\Phi _{\lambda \mu} ((\lambda
- \mu) \cdot x,z)$ when $\lambda\cdot\mu=-1$, that is, when $\mu=-\lambda$.
Denoting by $\alpha = \lambda$ the corresponding short root of $F_4$, the entry
in the Lax pair is given by the function $\Lambda (2 u,z)$ for $u=\a \cdot x =
\12 \a ^\vee \cdot X + \omega _2 \dv l^\vee (\a ^\vee)  $. Its asymptotics is
read off directly from (4.11). Since $\dv l^\vee (\a ^\vee) \geq 1 -\dv$, we
have $u\to +\infty$ and $u-\omega _2 \to -\infty$, and so only the first term
on the r.h.s. of (4.11) contributes,
$$
C_{\lambda,\mu}\Phi_{\lambda \mu}(2\ax,z)
\to 
2M_{|\alpha|}c_{\lambda,-\lambda} e^{\omega _2 (\dv -\dv l^\vee (\a ^\vee)) - 
\12 
\a ^\vee
\cdot X}.
\eqno (4.29)
$$
In the limit, only contributions from simple roots $\alpha$ satisfying $l^\vee
(\a ^\vee)=1$ remain, and we recover (4.4). 

\medskip

It remains to discuss the contributions arising from short positive roots when
$\lambda \cdot \mu=\12$. They occur under the form $\Phi _1 (u,z)$, for $u
=\half \a ^\vee \cdot X + \omega _2 \dv l^\vee (\a ^\vee )$. The limit of the
function requires some care, since the leading behavior as $\omega _2 \to
\infty$ cancels between the two terms in the definition of (4.27),
$$
\Phi _1 (u,z)
\to 
\mp 2 Z^{\mp 1} e^{\pm \12 u - \omega _2}
\qquad 
u \to \pm \infty.
\eqno (4.30)
$$
Since asymptotically, we have $u \sim \omega _2 \dv l^\vee (\a ^\vee)$, we see
that this contribution always converges to 0 since we always have $\half \omega
_2 \dv l^\vee (\a ^\vee) <1$. (Strictly speaking, we should also
check that the terms discarded when we approximated
$\theta_1^*({z\over 2\omega_1})$ and
$\theta_1^*({u\over 2\omega_1})$ by $2\,{\rm sinh}({z\over 2})$
and $2\,{\rm sinh}({u\over 2})$ do not contribute.
But this is also easily done.) Thus, the short roots $\alpha = \lambda - \mu$
with $\lambda \cdot \mu = \12$ do not survive in the Toda Lax pair, completing
the proof of (4.4) for this case.
Again, the evaluation of derivatives leading to (4.5) is a mere routine,
and the proof of Theorem 2 is complete.

\bigskip
\bigskip

\centerline{\bf ACKNOWLEDGEMENTS}

\bigskip

We would like to thank the Institute for Theoretical Physics in Santa Barbara 
for its hospitality during the January 1998 workshop on Geometry and Duality, 
where part of this work was done.

\bigskip
\bigskip

\centerline{\bf REFERENCES}

\bigskip

\item{[1]} D'Hoker, E. and D.H. Phong, ``Calogero-Moser Lax pairs with spectral 
parameters
for general Lie algebras", April 1998 preprint, hep-th/9804124.

\item{[2]} Donagi, R. and E. Witten, ``Supersymmetric Yang-Mills theory
and integrable systems", Nucl. Phys. {\bf B 460} (1996) 299-334,
hep-th/9510101.

\item{[3]} D'Hoker, E. and D.H. Phong,
``Calogero-Moser systems in SU($N$) Seiberg-Witten
theory", hep-th/9709053, Nucl. Phys. {\bf B513} (1998) 405.

\item{[4]} Martinec, E., ``Integrable structures in supersymmetric
gauge and string theory", Phys. Lett. {\bf B367} (1996) 91; hep-th/9510204.

\item{[5]} Krichever, I., unpublished communication.

\item{[6]} Inozemtsev, V.I., ``The finite Toda lattices",
Comm. Math. Phys. {\bf 121} (1989) 628-638.

\item{[7]} Inozemtsev, V.I., ``Lax representation with spectral parameter
on a torus for integrable particle systems",
Letters in Math. Physics {\bf 17} (1989) 11-17.

\item{[8]} D'Hoker, E. and D.H. Phong, ``Spectral Curves for 
Super-Yang-Mills with Adjoint Hypermultiplet for General Lie Algebras", April 
1998, hep-th/9804126.

\item{[9]} Martinec, E. and N. Warner, ``Integrable models and
supersymmetric gauge theory", Nucl. Phys. {\bf B 459} (1996) 97-112,
hep-th/9609161.

\item{[10]} Kac, V., {\it Infinite-dimensional Lie algebras},
Birkh\"auser, (1983) Boston.

\item{[11]} Goddard, P. and D. Olive, ``Kac-Moody and Virasoro algebras
in relation to quantum physics", Int. J. Mod. Phys. A,
Vol. {\bf I} (1986) 303-414.

\item{[12]} MacKay, W.G. and J. Patera, and D.W. Rand, {\it Tables of 
Representations of Simple Lie Algebras}, Vol I, ``Exceptional Simple Lie 
Algebras", Centre de Recherches Math\'ematiques Publications, Universit\'e de 
Montr\'eal, 1990.

\item{[13]} Olshanetsky, M.A. and A. Perelomov, ``Classical integrable
finite-dimensional systems related to Lie algebras",
Phys. Reports {\bf 71} (1981) 313-400;\hfil\break
Perelomov, A., {\it Integrable systems of classical mechanics 
and Lie algebras}, Birkh\"auser (1990) Boston;\hfil\break
Leznov, A.N.,  and M.V. Saveliev, {\it Group Theoretic Methods for Integration 
of Non-linear 
Dynamical Systems}, Birkhauser 1992; \hfil\break
M. Adler and P. van Moerbeke, ``Completely integrable systems,
Euclidian Lie algebras, and curves", Advances in Math. {\bf 38} (1980) 267-317;
\hfil\break
M. Adler and P. van Moerbeke, ``Linearization of Hamiltonian systems,
Jacobi varieties, and representation theory", Advances in Math. {\bf 38} (1980) 
318-379;
\hfil\break
M. Adler and P. van Moerbeke, ``The Toda lattice,
Dynkin diagrams, singularities and Abelian varieties",
Inventiones Math. {\bf 103} (1991) 223-278.

\item{[14]} Erdelyi, A., ed., {\it Higher Transcendental Functions}, Vol II, 
R.E. Krieger Publishing Co., Florida (1981).

\item{[15]} Krichever, I., ``Elliptic solutions of the Kadomtsev-Petviashvili
equation and integrable systems of particles",
Funct. Anal. Appl. {\bf 14} (1980) 282-290.

\item{[16]} Seiberg, N. and E. Witten,
``Electric-magnetic duality, monopole condensation, and
confinement in N=2 supersymmetric Yang-Mills theory",
Nucl. Phys. {\bf B 426} (1994) 19, hep-th/9407087.\hfil\break
Seiberg, N. and E. Witten,
``Monopoles, duality, and chiral symmetry breaking
in N=2 supersymmetric QCD", Nucl. Phys. {\bf B 431} (1994) 494,
hep-th/9410167.

\item{[17]} Argyres, P.C. and A.E. Faraggi, ``The Vacuum Structure and Spectrum 
of N=2 Supersymmetric SU(N) Gauge Theory", Phys. Rev. Lett. {\bf 74} (1995)
3931, hep-th/9411057.

\item{[18]} Klemm, A., W. Lerche, S. Theisen, and S. Yankielowicz, 
``Simple singularities and N=2 supersymmetric Yang-Mills theory",
Phys. Lett. {\bf B 344} (1995) 169, hep-th/9411048.

\item{[19]} Lerche, W., ``Introduction to Seiberg-Witten theory and its
stringy origins", Nucl. Phys. Proc. Suppl. {\bf 55B} (1997) 83, hep-th/9611190.

\item{[20]} D'Hoker, E., I.M. Krichever, and D.H. Phong,
``The effective prepotential for N=2 supersymmetric
SU($N_c$) gauge theories", Nucl. Phys. {\bf B 489} (1997) 179-210,
hep-th/9609041;
\hfill\break 
D'Hoker, E., I.M. Krichever, and D.H. Phong, ``The effective prepotential
for N=2 supersymmetric SO($N_c$) and Sp($N_c$) gauge theories",
Nucl. Phys. {\bf B 489} (1997) 211-222, hep-th/9609145;
\hfill\break
D'Hoker, E., I.M. Krichever, and D.H. Phong,
``The renormalization group equation for N=2 supersymmetric
gauge theories", Nucl. Phys. {\bf B 494} (1997), 89-104, hep-th/9610156.
\hfill\break
D'Hoker, E. and D.H. Phong, ``Strong coupling expansions in $SU(N)$
Seiberg-Witten theory", 
Phys. Lett. {\bf B 397} (1997) 94-103, hep-th/9701151.

\end